%% file: main.tex
\documentclass{article}

\usepackage[a4paper, total={6in, 10in}]{geometry}

\usepackage{multirow}
\usepackage{multicol}
\usepackage{graphicx}
\usepackage[utf8]{inputenc}
\usepackage{amsmath}
\usepackage{adjustbox}
\usepackage{todonotes}
\usepackage{caption} 
\usepackage{rotating}
\usepackage{xcolor}
\usepackage{hyperref}
\usepackage{verbatim}

\usepackage{float}

\usepackage{setspace} % double spacing
\usepackage{lineno} % line numbering

\usepackage[backend = biber,
style = numeric-comp,
giveninits = true,  
terseinits = true,
maxbibnames = 6,
minbibnames = 3,
sorting = none,
urldate=long,     % Prints "15 January 1999"
dateabbrev=true,  % Shortens January to Jan
isbn = false,
doi = false
] {biblatex}

\DeclareFieldFormat{urldate}{%
  Accessed\space
  \thefield{urlday}\space
  \mkbibmonth{\thefield{urlmonth}}\space
  \thefield{urlyear}}

\DefineBibliographyExtras{english}{%
  \protected\def\mkbibmonth#1{%
    \ifcase#1\or Jan\or Feb\or Mar\or Apr\or May\or June\or
    July\or Aug\or Sept\or Oct\or Nov\or Dec\fi}%
}

\DeclareFieldFormat{urldate}{Accessed\space\thefield{urlday}\space\mkbibmonth{\thefield{urlmonth}}\space\thefield{urlyear}}

\renewbibmacro*{urldate}{\printurldate}

\renewbibmacro{in:}{} % remove "In" in front of journal
\DeclareNameAlias{default}{family-given}
\DeclareFieldFormat[article]{title}{#1}

\DeclareFieldFormat[article]{journaltitle}{#1}
\DeclareFieldFormat[article]{pages}{#1}
\DeclareBibliographyDriver{article}{%
  \usebibmacro{author}%
  \setunit{\adddot\space}\newblock
  \usebibmacro{title}%
  \setunit{\adddot\space}\newblock
  \printfield{journaltitle}%
  \setunit{\adddot\space}%
  \printfield{year}%
  \setunit{\addsemicolon}%
  \printfield{volume}%
  \setunit{\addcolon}%
  \printfield{pages}%
  \usebibmacro{finentry}
}

\DeclareFieldFormat{url}{\url{#1}}
\DeclareFieldFormat[misc,online]{title}{#1}

\DeclareBibliographyDriver{online}{%
  \usebibmacro{author/editor+others}%
  \setunit{\adddot\space}\newblock
  \usebibmacro{title}%
  \setunit{\adddot\space}\newblock
  \printdate
  \setunit{\adddot\space}\newblock
  \printfield{note}% 
  \setunit{\adddot\space}\newblock
  \printfield{url}%
  \setunit{\adddot\space}\newblock
  \usebibmacro{urldate}%
  \usebibmacro{finentry}
}

\title{\bf{Adjusting for Outcome Reporting Bias in Meta-analysis: A Multiple Imputation Approach}}
\author{Cora Burgwinkel, Saverio Fontana and Leonhard Held\\ 
  Epidemiology, Biostatistics
  and Prevention Institute (EBPI)\\ 
  and Center for Reproducible Science and Research Synthesis (CRS) \\
  University of Zurich (UZH)\\ Hirschengraben 84,
  8001 Zurich, Switzerland\\ Corresponding author: \texttt{cora.burgwinkel@uzh.ch}}

\date{}

\input{newCommands}

\begin{document}
\begin{spacing}{1.2}

\maketitle
%\linenumbers
\section*{Abstract}

\textbf{Background:} Outcome reporting bias (ORB) occurs when study outcomes are selectively reported based on their results. ORB potentially undermines the credibility and validity of meta-analyses and contributes to research waste by distorting overall treatment effects. ORB can be viewed as a missing data problem in which unreported study outcomes introduce bias. Despite the serious implications ORB poses, it remains an underrecognized issue, with only a few adjustment methods available.\\[.25cm]
\noindent
\textbf{Methods:} We propose an approach that addresses unreported study outcomes in meta-analyses through multiple imputation for univariate and multivariate meta-analysis. To assess the impact of ORB in meta-analyses, we apply our proposed methodology to real clinical data affected by ORB, and conduct a simulation study to evaluate the method's performance under a range of scenarios.\\[.25cm]
\noindent
\textbf{Results:} The proposed method provides bias-adjusted estimates under assumed selective non-reporting mechanisms. In the application to clinical data, ORB-adjusted estimates were systematically shifted towards less extreme treatment effects compared with naive analyses, highlighting the potential magnitude of ORB in practice. The simulation study shows that the extent of adjustment depends on the assumed selection mechanism and the degree of heterogeneity, with stronger selection leading to larger adjustment.\\[.25cm]
\noindent
\textbf{Conclusions:} Imputing unreported study outcomes provides a promising approach to address ORB in meta-analyses. The multivariate approach extends ORB adjustment to jointly model correlated outcomes, allowing borrowing of strength across outcomes. Overall, we propose a practical and flexible approach for evaluating the sensitivity of univariate and multivariate meta-analytic conclusions to ORB.\\[.25cm]
\noindent
\textbf{Keywords:} Outcome Reporting Bias, Meta-analysis, Multiple Imputation, Selection Model

\newpage

\section{Background} \label{Introduction}
Selective reporting of outcomes in clinical trial reports refers to reporting only a subset of the originally recorded outcome variables in the final report. Selective outcome reporting can create outcome reporting bias (ORB) when the decision to report results within published studies is influenced by the significance or direction of the results \cite{chan2005identifying, lemmens2024outcome}. Previous reviews show that statistically significant outcomes are more likely to be fully reported than non-significant outcomes \cite{chan2005identifying, chan2014increasing, jones2015comparison, glasziou2014reducing, dwan2013systematic} and studies with significant results tended to be published earlier than studies with non-significant results \cite{song2010dissemination, williamson2005outcome}.

Published trial reports are expected to provide a complete and reliable account of study findings. Nevertheless, ORB is common \cite{chan2004outcome, beurden2021selective, ward2022outcome} and numerous forms of selective reporting have been documented in randomized controlled trials (RCTs) \cite{page2013many}. ORB threatens the validity of meta-analysis of both benefit \cite{kirkham2010impact} and harm \cite{saini2014selective} outcomes, because omitting outcomes on the basis of their results leads to meta-analyses being based on an incomplete and potentially biased subset of the available evidence. Beyond distorting treatment effect estimates, ORB can mislead clinicians and patients and contribute to research waste by encouraging unnecessary duplication of studies investigating outcomes that have already been measured but not reported \cite{chan2014increasing, glasziou2014reducing, thomas2022catalogue, butcher2020outcome}. ORB has been documented across a wide range of clinical areas, including psychiatry \cite{lancee2017outcome, shinohara2015protocol}, rehabilitation \cite{wang2023has}, healthcare and behavioural interventions \cite{page2014bias, matvienko2024selective, lancee2022selective, milette2011transparency} and surgical RCTs \cite{komukai2024publication}. Consequently, ORB undermines both the credibility of clinical evidence \cite{ioannidis2014clinical, frosi2015multivariate, littell2023protocol} and the transparency of clinical trial reporting \cite{souza2023selective}.
Although prospective trial registration is associated with a lower risk of reporting bias \cite{silva2024many, chan2017association, zhang2025threat}, mandatory trial registration has not prevented selective outcome reporting. Discrepancies between registered protocols and published reports remain common, even among trials published in leading medical journals \cite{jones2015comparison, howard2017systematic}. For example, a PLoS One systematic review of 137 RCTs from five high-impact medical journals (\textit{The Lancet, British Medical Journal, New England Journal of Medicine, Annals of Internal Medicine and Journal of American Medical Association}) found that 18\% (25/137) of the RCTs  had discrepancies related to the primary outcome \cite{fleming2015outcome}. To facilitate the identification of ORB in evidence synthesis, Dawn et al. \cite{dwan2010assessing} developed practical guidance for assessing selective outcome reporting while Page et al. \cite{page2023rob} introduced the ROB-ME (Risk Of Bias due to Missing Evidence) tool to assess the risk of bias arising from missing evidence in meta-analyses.  

Selective reporting is essentially a missing data problem and due to the lack of awareness of ORB, the common practice is to simply ignore the missing outcome data and carry out a univariate meta-analysis. In univariate meta-analyses the outcomes are analyzed separately, using complete case data, so only studies which report the outcome are included in the meta-analysis \cite{riley2007evaluation}. However, given ORB, studies that selectively omit or change outcomes of interest might distort the overall treatment effect because the analysis is based on a biased subset of the evidence \cite{thomas2022catalogue, dwan2010assessing}. In practice, systematic reviews often consider multiple correlated outcomes because each patient contributes data to several endpoints, leading to correlated effect estimates within studies. These outcomes can be jointly synthesized using multivariate meta-analysis, which accounts for their within-study correlation \cite{riley2007evaluation, jackson2011multivariate}. The idea is to borrow strength across outcomes, so to learn about the unreported study outcomes through the reported study outcomes \cite{kirkham2012multivariate}. However, the study-specific within-study correlation is rarely reported in practice and must often be approximated \cite{riley2009multivariate}. To address this, uncertainty in the correlation structure can be explored through sensitivity analyses. Compared with analysing each outcome separately, multivariate meta-analysis can improve estimation efficiency by incorporating information from correlated outcomes, particularly when some studies missing data for several outcomes \cite{jackson2011multivariate, riley2009multivariate, frosi2015multivariate}.

Even though ORB is less explored than publication bias, which describes the non-publication of studies due to lack of significance of the results, there are currently a few different statistical methodologies available to adjust for ORB in meta-analysis. Williamson and Gamble \cite{williamson2007application} suggested a maximum bias bound for ORB to assess the robustness of within-study selective non-reporting. Kirkham et al. \cite{kirkham2012multivariate} introduced a bivariate meta-analysis adjustment of two correlated outcomes and Bai et al. \cite{bai2021bayesian} proposed a Bayesian extension of it. Hwang and DeSantis \cite{hwang2018multivariate} suggested a Bayesian multivariate network meta-analysis to adjust for ORB and Liu et al. \cite{liu2018bayesian} investigated a Bayesian mixed treatment comparison meta-analysis for correlated outcomes. van Aert and Wicherts \cite{van2024correcting} suggested a meta-regression approach and Saracini and Held \cite{saracini2025addressing} proposed to address ORB through a selection model formulation. Among the available methods, the most widely used method for ORB-adjustment was developed by Copas et al. \cite{copas2019model}. The Copas adjustment method categorizes unreported outcomes into three bias risk categories: no risk, low risk, and high risk, according to the Outcome Reporting Bias in Trials (ORBIT) classification system. Based on this classification, Copas et al. \cite{copas2019model} introduced a likelihood-based ORB-adjustment method and then the adjustment is applied separately to each outcome in the meta-analysis. The Copas adjustment method assumes that treatment effects, and potentially standard errors, are unreported, whereas study sample sizes are reported.

In the following, a meta-analysis is considered, and the missing event frequencies are imputed and then combined with importance sampling. The idea builds on the work by Carpenter et al. \cite{carpenter2011assessing}, who proposed a multiple imputation approach for selection bias: unreported studies are first imputed under a missing at random (MAR) assumption, and then reweighted according to a missing not at random (MNAR) selection mechanism. A similar problem was previously addressed by Williamson and Gamble \cite{williamson2005identification}, who explored the impact of outcome selection bias by imputing unreported outcomes under a range of deterministic assumptions, without explicitly modeling the selection process. In contrast, the approach proposed in this work integrates multiple imputation with probabilistic selection via importance sampling, and applies it to address ORB in both univariate and multivariate meta-analyses. The effectiveness of the introduced ORB-adjustment is investigated in both univariate and multivariate meta-analysis models. As for the Copas selection model \cite{copas2019model}, we assume that study sample sizes are reported while treatment effects and standard errors are selectively unreported. 

This paper is structured as follows: Section \ref{Methods} outlines the multiple imputation and weighting approach employed for addressing ORB in meta-analysis. Section \ref{application} details the application to a real clinical data set for both the univariate and multivariate approach. In Section \ref{SimulationStudy}, a simulation study that examines the effects of ORB and evaluates the effectiveness of the proposed ORB-adjustment method within the context of a random-effects meta-analysis model is presented. Section \ref{discussion} provides a discussion summarizing the methodology and findings, along with their limitations and finally, Section \ref{conclusion} concludes this work.

\section{Methods ORB-adjustment} \label{Methods}

In this section, we present the methodology used to adjust for ORB by imputing unreported study outcomes in meta-analysis. We first introduce the univariate and multivariate meta-analysis models, followed by the selection models and the importance sampling approach underlying the proposed adjustment method.

\subsection{Overall Summary Estimate}
The observed study-specific effect estimates $\hat\theta_i$ and their corresponding standard errors $\sigma_i$ were calculated from $2 \times 2$ contingency tables from study $i = 1, \dots, K$ using either the log odds ratio (OR) or the log relative risk (RR). In studies with one or more zero cell counts, a continuity correction was applied following Agresti \cite[p. 71]{agresti2013categorical}. These study-specific effect estimates are combined to an overall summary estimate by calculating a weighted average across studies. In the following, first the univariate meta-analysis is introduced, which analyses outcomes separately and then the multivariate meta-analysis, which considers multiple outcomes simultaneously. 

\subsubsection{Univariate Random-Effects Meta-Analysis} \label{REM}
Throughout this work, we consider a random-effects model (REM) which assumes that there are different true effects among the studies due to heterogeneity, and accounts for this by including random effects for each study. The REM for a single beneficial outcome is
\begin{center}
$\hat\theta_i \sim \Nor(\theta + \nu_i, \sigma^{2}_i)$
\end{center}
where $\theta$ represents the overall average treatment effect across studies, $\nu_i$
is the deviation of $i$-th study effect to $\theta$, and $\sigma_i$ is the within-study standard deviation. Additionally, the deviations $\nu_i$ are assumed to follow a normal distribution:
$\nu_i \sim \Nor(0, \tau^2)$.
Here, $\tau^2$ denotes the between-study heterogeneity variance which is estimated from the data. The random-effects meta-analytic (MA) estimate of the treatment effect is 
\begin{center}
$\hat\theta_{\MA, \mbox{\tiny{RE}}} = \frac{\sum_{i \in \Rep} w_i \hat \theta_i}{\sum_{i \in \Rep} w_i}$ with weights $w_i = \frac{1}{\sigma^{2}_i + \tau^2}$.
\end{center}

The choice of these weights reflects both within-study variability and between-study heterogeneity, which minimizes the uncertainty of $\hat\theta_{\MA, \mbox{\tiny{RE}}}$.  The standard error is $\SE(\hat\theta_{\MA, \mbox{\tiny{RE}}}) = 1/{\sqrt{\sum_{i \in \Rep} w_i}}.$ 

\subsubsection{Multivariate Random-Effects Meta-Analysis}
\label{multivariate}
A multivariate meta-analysis jointly synthesizes multiple outcomes across studies and provides an alternative to performing separate univariate meta-analyses for each outcome. The outcomes are often correlated because each patient contributes data to multiple endpoints. In the presence of ORB, where not all studies report all outcomes, the multivariate approach allows borrowing of strength across correlated outcomes \cite{jackson2011multivariate, kirkham2012multivariate, schmid2020handbook}. The idea is to learn about the unreported study outcomes through the reported (correlated) study outcomes. We focus on the bivariate REM with two outcomes. Let $i = 1,\dots,K$ index the studies and $j \in \{1,2\}$ the outcomes. The study-specific effect estimate vector is assumed to follow
\begin{equation*}
\label{eq:multivariate_hetero}
\widehat{\boldsymbol{\theta}}_{i} =
\begin{pmatrix}
\hat{\theta}_{i1}\\
\hat{\theta}_{i2}
\end{pmatrix}
\sim
\Nor\left(
\begin{pmatrix}
\theta_1\\
\theta_2
\end{pmatrix},
\boldsymbol{\Sigma}_i+\boldsymbol{\Psi}
\right),
\end{equation*}
where $\boldsymbol{\Sigma}_i$ is the within-study covariance matrix and $\boldsymbol{\Psi}$ is the between-study variance-covariance matrix:
\begin{equation*}
\boldsymbol{\Sigma}_i=
\begin{pmatrix}
\sigma_{i1}^2 & \rho_W \sigma_{i1}\sigma_{i2}\\
\rho_W \sigma_{i1}\sigma_{i2} & \sigma_{i2}^2
\end{pmatrix} \text{and }
\boldsymbol{\Psi}=
\begin{pmatrix}
\tau_1^2 & \rho_B\tau_1\tau_2\\
\rho_B\tau_1\tau_2 & \tau_2^2
\end{pmatrix}.
\end{equation*}
$\rho_W$ and $\rho_B$ denote the within- and between-study correlations, respectively, and in practice, $\rho_W$ is rarely reported and therefore often has to be approximated \cite{riley2009multivariate, jackson2011multivariate}. A common approach is to estimate $\rho_W$ using the Pearson correlation calculated from studies reporting both outcomes \cite{kirkham2012multivariate}. Alternatively, sensitivity analyses over a plausible range of correlations or external information, such as individual participant data (IPD), can be used to narrow the range of likely values \cite{jackson2011multivariate, kirkham2012multivariate, mavridis2013practical, schmid2020handbook}. The bivariate REM forms the basis of the proposed multivariate ORB-adjustment method by accounting for both within-study and between-study variability while allowing information to be borrowed across correlated outcomes.

\subsection{Imputation of Missing Standard Errors} \label{ImputationSE}

If the total sample sizes $n_i = n_{ti} + n_{ci}$ are available for all the studies, it is possible to impute unreported standard errors $\tilde\sigma_{ij}$ for study $i$ and outcome $j$. Knowledge of the standard error of the unreported study outcomes is required to be able to calculate the MA estimate based on all studies, also on the unreported study outcomes. The imputation is performed separately for each outcome. We impute missing standard errors for an unreported study $i$ and outcome $j$ with \cite{copas2014model}
\begin{equation} \label{eq:missingVar}
\tilde\sigma_{ij} = \frac{1}{\sqrt{\hat k_j \, n_i}} \mbox{ where } \hat k_j  = \frac{\sum_{i \in \Rep_j} \sigma_{ij} ^{-2}}{\sum_{i \in \Rep_j} n_i}
\end{equation}
where $\text{Rep}_j$ denotes the studies reporting outcome $j$.

\subsection{Imputation via Multivariate Normal Distribution}
\label{ImputationMVN}

When some studies do not report the study outcome of interest, a multivariate normal distribution can be used to impute the unreported study outcomes conditional on the reported study outcomes. The multivariate normal distribution allows the modeling of multiple variables that may be correlated with each other. The key advantage of the multivariate normal model is that conditional distributions of subsets (\eg the unreported studies) are also normally distributed. With the multivariate normal distribution all unreported study outcomes can be imputed simultaneously under a MAR assumption. 

\subsubsection{Univariate Approach} 
\label{univariateMVN}

Let $\hat{\theta}_{j,\MA}$ denote the naive univariate MA estimate obtained from the reported studies for a single outcome $j$. In the univariate case, we define the complete vector of underlying study-specific effects across all $K$ studies for outcome $j$ as $\widehat{\boldsymbol{\theta}}_j = (\hat{\theta}_{1j}, \dots, \hat{\theta}_{Kj})^\top$. To impute the unreported study outcomes under a MAR assumption, we assume this joint vector follows a multivariate normal distribution centered on the naive MA estimate

\begin{equation*}
\widehat{\boldsymbol{\theta}}_j \sim \Nor\left( \boldsymbol{1}_K \hat{\theta}_{j,\MA}, \boldsymbol{\Sigma}_j \right),
\end{equation*}
where $\boldsymbol{1}_K$ is a column vector of ones of length $K$. The total covariance matrix $\boldsymbol{\Sigma}_j \in \mathbb{R}^{K \times K}$ is defined as

\begin{equation*} \label{eq:sigma_uni}
\boldsymbol{\Sigma}_j = \diag (\sigma_{1j}^2 + \tau_j^2, \dots, \sigma_{Kj}^2 + \tau_j^2) + \SE(\hat\theta_{j, \MA})^2 \boldsymbol{J}_K
\end{equation*}
where $\sigma_{ij}^2$ represents the within-study variances, including both the reported variances and unreported variances (imputed following \eqref{eq:missingVar}), $\tau_j^2$ is the between-study variance estimated from the reported studies, 
$\SE(\hat\theta_{j, \MA})$ is the standard error of the naive MA estimate, and $\boldsymbol{J}_K$ denotes the 
$K \times K$ matrix of ones. The complete vector of underlying study-specific effects $\widehat{\boldsymbol{\theta}}_j$ and its corresponding total covariance matrix $ \boldsymbol{\Sigma}_j$ are ordered such that the $K_R$ reported studies appear first and the $K_U$ unreported studies follow. Therefore, both the vector and the matrix can be partitioned into reported (R) and unreported (U) components

\begin{equation} \label{eq:partitioned_uni}
\widehat{\boldsymbol{\theta}}_j = 
\begin{pmatrix}
\widehat{\boldsymbol{\theta}}_{j,R} \\
\widehat{\boldsymbol{\theta}}_{j,U}
\end{pmatrix}, 
\qquad
\boldsymbol{\Sigma}_j = 
\begin{pmatrix}
\boldsymbol{\Sigma}_{j,RR} & 
\boldsymbol{\Sigma}_{j,RU} \\
\boldsymbol{\Sigma}_{j,UR} 
& \boldsymbol{\Sigma}_{j,UU}
\end{pmatrix}.
\end{equation}
Conditioning on the reported study outcomes $\widehat{\boldsymbol{\theta}}_{j,R}$, the conditional distribution of the unreported study outcomes $\widehat{\boldsymbol{\theta}}_{j,U}$ is \cite[p. 349]{held2020likelihood}
\begin{align}
\widehat{\boldsymbol{\theta}}_{j,U} \given \widehat{\boldsymbol{\theta}}_{j,R} 
&\sim \Nor(\boldsymbol{\theta}_{j,U \given R},\, \boldsymbol{\Sigma}_{j,U \given R}), \quad \text{where} \label{eq:condDist} \\
\boldsymbol{\theta}_{j,U \given R} 
&= \boldsymbol{1}_{K_U} \hat{\theta}_{j,\MA} + \boldsymbol{\Sigma}_{j,UR} \left( \boldsymbol{\Sigma}_{j,RR} \right)^{-1} \left( \widehat{\boldsymbol{\theta}}_{j,R} - \boldsymbol{1}_{K_R} \hat{\theta}_{j,\MA} \right) \quad \text{and} \nonumber \\
\boldsymbol{\Sigma}_{j,U \given R} 
&= \boldsymbol{\Sigma}_{j,UU} - \boldsymbol{\Sigma}_{j,UR} \left( \boldsymbol{\Sigma}_{j,RR} \right)^{-1} \boldsymbol{\Sigma}_{j,RU}. \nonumber
\end{align}
In order to impute the effect estimates of the unreported study outcomes, we sample from the conditional distribution \eqref{eq:condDist} for $m = 1, \dots, M$ where $M$ is the number of imputations.

\subsubsection{Bivariate Approach} 
\label{bivariateMVN}

The bivariate approach extends the multivariate normal framework to jointly impute unreported study outcomes across two correlated outcomes. We define the complete vector of underlying study-specific effects in a study-major format (grouped by study) as $\widehat{\boldsymbol{\theta}} = (\hat{\theta}_{11}, \hat{\theta}_{12}, \dots, \hat{\theta}_{K1}, \hat{\theta}_{K2})^\top$ where $\hat{\theta}_{i1}$ and $\hat{\theta}_{i2}$ denote the observed effects for outcomes 1 and 2 in study $i$. Then, under the same normality assumption used in the univariate case, we can write
\begin{equation*}
    \widehat{\boldsymbol{\theta}} 
\sim \Nor\left( \boldsymbol{1}_K \otimes \widehat{\boldsymbol{\theta}}_{\MA}, \boldsymbol{\Sigma} \right)
\end{equation*}
where $\widehat{\boldsymbol{\theta}}_{\MA} = (\hat{\theta}_{1,\MA}, \hat{\theta}_{2,\MA})^\top$ represents the vector of naive bivariate MA estimates derived from the reported study outcomes and $\boldsymbol{1}_K$ is a column vector of ones of length $K$.
The total covariance matrix $\boldsymbol{\Sigma} \in \mathbb{R}^{2K \times 2K}$ is constructed as

\begin{equation*} 
\label{eq:sigma_multi_rem} \boldsymbol{\Sigma} = \boldsymbol{V} + (\boldsymbol{I}_K \otimes \boldsymbol{\Psi}) + (\boldsymbol{J}_K \otimes \Cov(\widehat{\boldsymbol{\theta}}_{\MA})). 
\end{equation*}

The within-study covariance matrix $\boldsymbol{V} \in \mathbb{R}^{2K \times 2K}$ is a block-diagonal matrix that combines the study-specific covariance matrices $\boldsymbol{V}_i$ along its diagonal: $\boldsymbol{V} = \operatorname{bdiag}(\boldsymbol{V}_1, \dots, \boldsymbol{V}_K)$. Each block $\boldsymbol{V}_i$ represents the within-study covariance for study $i$

\begin{equation*} 
\boldsymbol{V}_i = 
 \begin{pmatrix}
\sigma_{i1}^2 & \rho_W \, \sigma_{i1} \sigma_{i2} \\
\rho_W \, \sigma_{i1} \sigma_{i2} & \sigma_{i2}^2
\end{pmatrix} 
\end{equation*}
where $\sigma_{i1}^2$ and $\sigma_{i2}^2$ are the within-study variances for each outcome, taking both reported and unreported study outcomes (imputed for each outcome separately following \eqref{eq:missingVar}) into account. The within-study correlation $\rho_W$ is estimated with the Pearson correlation on all reported study outcomes. $\boldsymbol{\Psi}$ is the $2 \times 2$ between-study variance-covariance matrix estimated on the reported studies and $\boldsymbol{I}_K$ is the $K \times K$ identity matrix.
$\Cov(\widehat{\boldsymbol\theta}_{\MA})$ is the variance-covariance matrix of the MA estimates extracted from the naive bivariate model. By incorporating the off-diagonal covariance term, the imputation model borrows information from the reported study outcomes from both outcomes. To exploit the properties of the conditional multivariate normal for imputation,  $\widehat{\boldsymbol{\theta}}$ and its corresponding total covariance matrix $\boldsymbol{\Sigma}$ are reordered from  study-major to outcome-major format to facilitate the partitioning into reported and unreported components following \eqref{eq:partitioned_uni}. The unreported studies are then imputed from the conditional distribution as in \eqref{eq:condDist}. In the bivariate case $\boldsymbol{\widehat\theta}_{j,\MA}$ is replaced with $\boldsymbol{\widehat\theta}_{\MA}$.

The main difference between the univariate and bivariate approach is that in the univariate case, unreported studies are imputed separately from the conditional distribution for each outcome, while in the bivariate case, the correlation between the outcomes is considered to impute unreported study outcomes for both outcomes simultaneously from the conditional distribution.

\subsection{Selection Models and Importance Sampling} \label{Importance}
Importance sampling is used to combine the simulated imputations of unreported study outcomes and weight the imputed data to yield an unbiased estimate \cite{tabandeh2022review}. Selection models explicitly account for the assumed missing data mechanism to correct for bias \cite{cooper2019handbook, hedges1992modeling}.  Carpenter et al. \cite{carpenter2011assessing} proposed a logistic selection model, which defines the probability of a study being reported, in our case, the probability of a study outcome being reported, as

\begin{equation} \label{eq:selection}
\Pr\mbox{(reporting outcome $j$ in study $i$)} = \text{expit} \left( \alpha + \delta \cdot \hat\theta_{ij} \right)
\end{equation}
where $\text{expit}(x) = 1/(1+\exp(-x))$, $\delta$ is the selection weight and $\hat\theta_{ij}$ is the estimated summary statistic for study $i$ and outcome $j$. The selection weight $\delta$ is often chosen over a range of values to assess the sensitivity of nonrandom selection. $\delta = 0$ corresponds to a special MAR case, implying that selection has no influence on inference. The logistic selection model above is common to both the univariate and bivariate ORB-adjustment, but the approaches differ in the construction of the importance weights. For the bivariate ORB-adjustment, the weight for imputations $m = 1, ..., M$, where $M$ is the total number of imputations, is 
\begin{equation} \label{eq:weights}
w^{(m)} \propto \exp \left( -\delta \sum_{j=1}^{2}  \sum_{i \in U_j} \hat\theta^{(m)}_{ij} \right)
\end{equation}
where $\hat\theta^{(m)}_{ij}$ is the imputed summary statistic from study $i$, outcome $j$ and imputation $m$. 
\eqref{eq:weights} assumes a common selection mechanism across outcomes, however, in principle, outcome-specific selection weights $\delta_j$ can also be considered. For the univariate ORB-adjustment, the summation over outcomes is omitted, yielding separate importance weights for each outcome. Note that the weights $w^{(m)}$ no longer depend on $\alpha$. They are subsequently normalized across all $M$ imputations such that $\sum_{m=1}^{M} w^{(m)} = 1$. Selection occurs on the effect estimates, meaning that studies with larger effect estimates are more likely to be reported. 
Regarding weight derivation, \eqref{eq:weights} is the only part which differs between successive imputations $m$ for the same study \cite[web appendix]{carpenter2011assessing}.  The adjusted estimate $\hat\theta_{\Adj}$ (for the current choice of $\delta$) is then 
\begin{equation} \label{eq:weightedAverage}
\hat\theta_{\Adj} = \sum^{M}_{m = 1} w^{(m)} \cdot \hat\theta^{(m)}_{\MA} 
\end{equation}
where $\hat\theta^{(m)}_{\MA}$ is the pooled MA estimate (or vector of estimates in the bivariate case) derived from the $m$-th completed dataset. The MA estimate $\hat\theta^{(m)}_{\MA}$ is based on the reported study outcomes and the unreported study outcomes imputed following Section \ref{ImputationMVN}. %Note that for the bivariate approach, these equations generalize to vectors and covariance matrices.

In comparison to the logistic selection model \eqref{eq:selection}, the selection models proposed by Saracini and Held \cite{saracini2025addressing} are based on the one-sided $p$-values $p$ with $\alpha$ as threshold for significance, \eg $\alpha = 0.05$, so the $p$-value is used to model the probability of selection. The selection models are often defined as a function of the $p$-value and are directly dependent on $\alpha$. For example, the simplest selection function for publication bias assumes that a study is published if $p \le \alpha$ \cite{hedges1984estimation, sutton2000modelling}. 

Selection based on one-sided $p$-values is related to selection based on the $z$-score, also explored by Carpenter et al. \cite{carpenter2011assessing}.
The $z$-scores are derived directly from the effect estimate and given as \cite[p.170]{egger2022systematic}
\begin{equation} \label{eq:z-score}
z_{ij}^{(m)} = \frac{\hat\theta^{(m)}_{ij}}{\SE(\hat\theta_{ij})}, \quad \text{for } i \in U_j
\end{equation}
where $\hat\theta^{(m)}_{ij}$ is the imputed effect estimate for study $i$, outcome $j$ and imputation $m$ and $\SE(\hat\theta_{ij})$ is the corresponding standard error (for the unreported study outcomes imputed following \eqref{eq:missingVar}). Selection on the $z$-score acts on the statistical significance, so studies with higher $z$-scores (and smaller $p$-values) are more likely to be reported. For selection on the $z$-score, the weights from \eqref{eq:weights} are calculated replacing $\hat\theta_{ij}^{(m)}$ with  $\hat z_{ij}^{(m)}$ from \eqref{eq:z-score}. Next, the meta-analysis model is calculated for each simulation $m$ based on the reported and imputed summary statistics, which gives the  MA estimate $\hat\theta^{(m)}_{\MA}$. Finally, the adjusted estimated $\hat\theta_{\Adj}$ is calculated following \eqref{eq:weightedAverage}.

The within- and between-imputation variances are
\begin{equation*}
\hat\sigma_{\mbox{\tiny{W, MNAR}}}^2 = \sum^M_{m=1} \hat\sigma_{\hat\theta_{\MA}}^{2, (m)} w^{(m)} \mbox{ and }
\hat\sigma_{\mbox{\tiny{B, MNAR}}}^2 = \sum^M_{m=1} w^{(m)} (\hat\theta^{(m)}_{\MA} - \hat\theta_{\Adj})^2, 
\end{equation*}
respectively. Following Rubin's rule the variance of the adjusted estimate is then
\begin{equation} \label{eq:var}
\Var(\hat\theta_{\Adj}) = \hat\sigma_{\mbox{\tiny{W, MNAR}}}^2 + \hat\sigma_{\mbox{\tiny{B, MNAR}}}^2.
\end{equation}
In summary, a model is fitted to the reported studies and then the unreported studies are imputed. Next, given $\delta$, the imputed data are reweighted under a nonrandom selection to get the adjusted estimate. 

\section{Application}
\label{application}

To evaluate the proposed ORB-adjustment, we apply it to the Cochrane systematic review \textit{Topiramate add-on for drug-resistant partial epilepsy} \cite{pulman2014topiramate}, which has previously been used by Copas et al. \cite{copas2019model}. We investigate how the ORB-adjustment differs for RR and OR and compare selection on the effect estimate to selection on the $z$-score. This is examined across a range of selection weights $\delta \in \{0.0, 0.1, 0.2, \ldots, 1.3\}$, where $\delta = 0.5$ represents moderate selection \cite{carpenter2011assessing}. 

The meta-analysis, originally conducted by Bresnahan et al. \cite{bresnahan2019topiramate}, includes studies examining the effect of Topiramate as an add-on treatment for drug-resistant focal epilepsy.  Twelve of the outcomes included in the meta-analysis were considered harmful and two outcomes beneficial. Since we focus on ORB-adjustment for beneficial outcomes, we only consider the two outcomes assumed to have a positive effect, \ie \textit{50\% seizure reduction} and \textit{seizure freedom}. \textit{50\% Seizure reduction} was reported for eleven of the twelve studies, while \textit{seizure freedom} was only reported in six of the twelve studies (see Table \ref{tab:data}). We decided to include study 12 (Coles, 1999) in the analyses to facilitate comparison between the univariate and multivariate ORB-adjustment. The data set contains study sample sizes and event counts for intervention and control groups. 

Effect sizes and standard errors are computed with a continuity correction of 0.5 applied for zero cells following recommendations in the literature \cite{copas2019model, agresti2013categorical}. Table \ref{tab:data} provides an overview of the log RR and the corresponding standard errors for each study. For the unreported studies (NA in Table \ref{tab:data}), the standard errors are imputed using \eqref{eq:missingVar}. The imputed summary statistics for the unreported studies are drawn from a multivariate normal distribution, described in Section \ref{ImputationMVN}. The process of drawing the unreported effect estimates from the multivariate normal distribution is repeated $M = 1000$ times, since a large number of imputations is recommended to ensure that high weights are distributed over a good number of imputations \cite{carpenter2011assessing}. Next, the logistic selection model is applied, and the simulated imputations are combined with the normalized weights from \eqref{eq:weights}. For selection on the $z$-score, the $z$-scores for each imputation $m$ are calculated following \eqref{eq:z-score}. Finally, based on the reported and imputed data, the adjusted estimate is derived as in \eqref{eq:weightedAverage} and its total variance is calculated using Rubin's rule as specified in \eqref{eq:var}. We first investigate the univariate ORB-adjustment (see Section \ref{univariateMVN}) and compare it to the bivariate approach (see Section \ref{bivariateMVN}).

\begin{table}[H]
\centering
\caption{Example meta-analysis data providing the event frequencies for beneficial outcomes \textit{50\% seizure reduction} and \textit{seizure freedom} among treated (T) and control (C) patients \cite{copas2019model, bresnahan2019topiramate}. The RR and its standard error are provided as well (for the unreported studies the standard errors are imputed following \eqref{eq:missingVar}).}
\label{tab:data}
\begin{adjustbox}{width=1\textwidth,center=\textwidth}
\begin{tabular}{ccccccccccccc}
\hline
\textbf{} &                   & \multicolumn{2}{l}{Sample Size} & \multicolumn{4}{c}{50\% Seizure Reduction}    & \multicolumn{4}{c}{Seizure Freedom}             \\ \hline \\[-0.4cm] 
\multicolumn{2}{c}{Study}     & T  & C   & T  & C  & $\hat\theta_i$  & $\sigma_i$ & T  & C   & $\hat\theta_i$   & $\sigma_i$ \\ \hline 
1  & Ben-Menachem 1996 & 28  & 28    & 12  & 0  & 3.22   & 1.42  & \textcolor{blue}{NA} & \textcolor{blue}{NA} & \textcolor{blue}{NA} & 1.61       \\2     & Elterman 1999   & 41   & 45  & 16  & 9  & 0.67  & 0.36    & 4   & 2  & 0.79  & 0.84       \\3  & Faught 1996 & 136 & 45  & 54   & 8   & 0.8    & 0.34 & \textcolor{blue}{NA} & \textcolor{blue}{NA} & \textcolor{blue}{NA} & 0.89 \\4  & Guberman 2002   & 171  & 92       & 77  & 22 & 0.63    & 0.20   & 10  & 2   & 0.99    & 0.76 \\5  & Korean 1999 & 91   & 86   & 45    & 11   & 1.35 & 0.30  & 7   & 1   & 1.89  & 1.06       \\6   & Privitera 1996    & 143 & 47  & 58    & 4  & 1.56  & 0.49   & \textcolor{blue}{NA} & \textcolor{blue}{NA} & \textcolor{blue}{NA}& 0.87       \\7  & Rosenfeld 1996 & 167 & 42 & 86  & 8  & 0.99   & 0.33  & \textcolor{blue}{NA} & \textcolor{blue}{NA} & \textcolor{blue}{NA} & 0.83 \\ 8   & Sharief 1996  & 23  & 24  & 8   & 2  & 1.43 & 0.73 & 2  & 0 & 1.65  & 1.52 \\9   & Tassinari 1996    & 30  & 30  & 14  & 3 & 1.54  & 0.58       & 0   & 0  & 0 & 1.98 \\10  & Yen 2000 & 23  & 23 & 11  & 3   & 1.29& 0.58  & \textcolor{blue}{NA} & \textcolor{blue}{NA} & \textcolor{blue}{NA} & 1.77 \\11 & Zhang 2011  & 46 & 40 & 22    & 3  & 1.85  & 0.58  & 0      & 0  & -0.14 & 1.99 \\12 & Coles 1999        & 52 & 51 & \textcolor{blue}{NA} & \textcolor{blue}{NA} & \textcolor{blue}{NA} & 0.42  & \textcolor{blue}{NA} & \textcolor{blue}{NA} & \textcolor{blue}{NA} & 1.86  \\ \hline 
\end{tabular}
\end{adjustbox}
\end{table}

The analysis compares ORB-adjusted estimates with naive MA estimates based only on reported studies. All figures focus on the outcome \textit{seizure freedom}, as it has more unreported study outcomes and provides a clearer visualization of our ORB-adjustment method. The corresponding figures for the outcome \textit{50\% seizure reduction} can be found in the supplementary material.

\subsection{Univariate ORB-adjustment}

Univariate ORB-adjustment is applied separately to both outcomes.  Even though heterogeneity is low ($I^2 = 8.9\%$ for \textit{50\% seizure reduction} and $I^2 = 0\%$ for \textit{seizure freedom}), we applied a REM as it is widely used in practice due to its more conservative estimation and ability to account for unmeasured heterogeneity \cite{higgins2002quantifying, higgins2009re}. We followed the Cochrane review \cite{pulman2014topiramate} by calculating 95\% confidence intervals (CIs) for benefit outcomes and implemented the univariate approach in R using \texttt{rma()} (equivalent to \texttt{rma.uni()}) from the \texttt{metafor} package.

\begin{figure}[!ht]
\centering
\caption{Forest plot for univariate and bivariate ORB-adjustment applied to Topiramate data for REM (with selection weight $\delta = 0.5$). For the bivariate analyses, $r$ denotes the assumed within-study correlation.}
\label{fig:Forestplot}
\includegraphics[width = 1\linewidth]{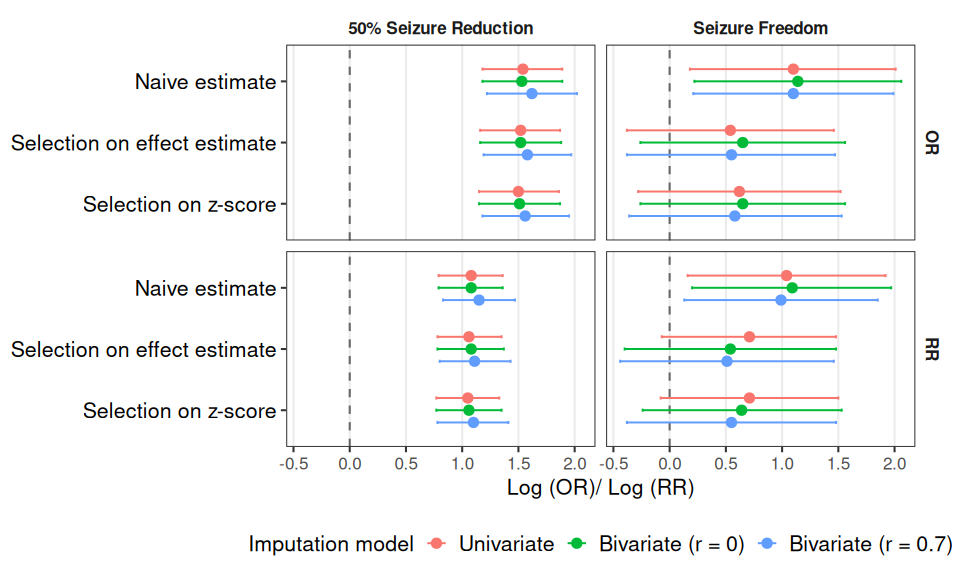}
\end{figure}

\begin{figure}[!ht]
\centering
\caption{Outcome \textit{seizure freedom}: comparison of naive and univariate ORB-adjusted estimates for selection on the log OR/log RR and on the $z$-score over increasing selection.}
\label{fig:OR_RR_REM}
\includegraphics[width = 1\linewidth]{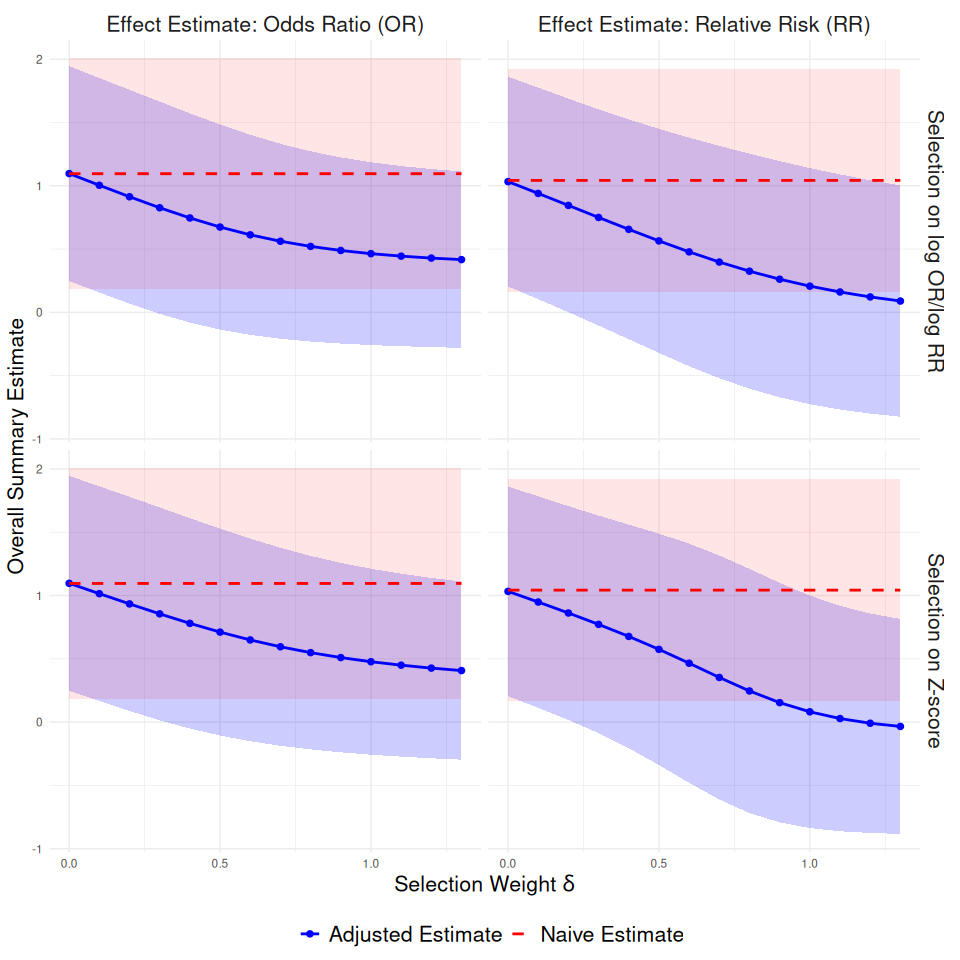}
\end{figure}

As shown in Figure \ref{fig:Forestplot}, adjustment for ORB had little impact on the estimated treatment effect for \textit{50\% seizure reduction}, with both the univariate and bivariate approaches producing estimates and 95\% CIs that were very similar to the naive estimates. This is expected as there is only one unreported study, and it is consistent with previous findings \cite{saracini2025addressing}. In contrast, for \textit{seizure freedom}, where half of the studies did not report outcome data, the ORB-adjusted estimates were consistently attenuated toward the null compared to the naive estimates, suggesting that ignoring ORB may overestimate the treatment effect when beneficial outcomes are selectively reported.
Figure \ref{fig:OR_RR_REM} shows how large the overestimation of the treatment effect is for \textit{seizure freedom} for increasing selection. Selection on the $z$-score results in slightly lower adjusted estimates than treatment effect selection. Differences between OR and RR are small but consistent, with OR yielding slightly larger estimates and wider CIs. With increasing $\delta$ (\ie $\delta > 0.8$), the adjusted OR decreases more sharply on the $z$-score selection since studies with larger standardized effects are preferentially retained, producing a stronger adjustment as the assumed selection strength increases.

\subsection{Bivariate ORB-Adjustment}
The bivariate ORB-adjustment simultaneously adjusts for both outcomes. Because the within-study correlations were unavailable, we approximate the correlation structure using the Pearson correlation between the observed study-level effect estimates among studies reporting both outcomes \cite{kirkham2012multivariate, jackson2011multivariate}. The estimated Pearson correlation between the observed study-level effect estimates is $r = -0.33$ (95\% CI: -0.90 to 0.66) for the log OR and $r = -0.31$ (95\% CI: -0.90 to 0.67) for the log RR. We use this quantity as an empirical proxy for the unknown within-study correlation in the multivariate imputation procedure and consider three approaches for incorporating the within-study correlation into the multivariate imputation procedure: (i) a single global correlation for all studies to capture uncertainty in the global correlation while keeping the correlation structure constant across studies. The correlation is modelled on the Fisher $z$-scale, $z = \arctanh(r)$, and we assume a normal distribution $z \sim \Nor(\mu_z, \sigma_z^2)$, where $\sigma_z^2$ denotes the variance of the Fisher z-transformed correlation. The mean, $\mu_z$, and variance, $\sigma_z^2$, are derived from the 95\% CI of the estimated Pearson correlation. A single value is drawn from this distribution and back transformed to the correlation scale. This sampled correlation is then used uniformly across all studies in the imputation step. (ii) a study-specific correlation to assume heterogeneity in the correlation structure across studies. The Fisher $z$-scale is used, but now a separate value $z_i \sim \Nor(\mu_z, \sigma_z^2)$ is drawn for each study $i$, and back transformed to obtain $r_i = \tanh(z_i)$ \cite[p. 104]{held2020likelihood}. This method allows the correlation to vary across studies while still being informed by the overall estimated uncertainty. (iii) fixed sensitivity values within the estimated CI, \ie $r \in \{-0.9, -0.6, -0.3, 0, 0.3, 0.6, 0.67\}$, thereby considering a low to very strong within-study correlation.
Multivariate models are fitted with a block-diagonal within-study covariance matrix supplied to the \texttt{rma.mv()} function via the V argument. For the random effects, we specified \texttt{random = $ \sim$ outcome | study} with \texttt{struct = "UN"}, which assumes an unstructured variance–covariance matrix for the study-level random effects.

\begin{figure}[ht!]
\centering
\caption{Outcome \textit{seizure freedom}: comparison of within-study correlations for bivariate ORB-adjustment for selection on the log OR/log RR and selection on the $z$-score over increasing selection.}
\label{fig:Multivariate1}
\includegraphics[width = 1\linewidth]{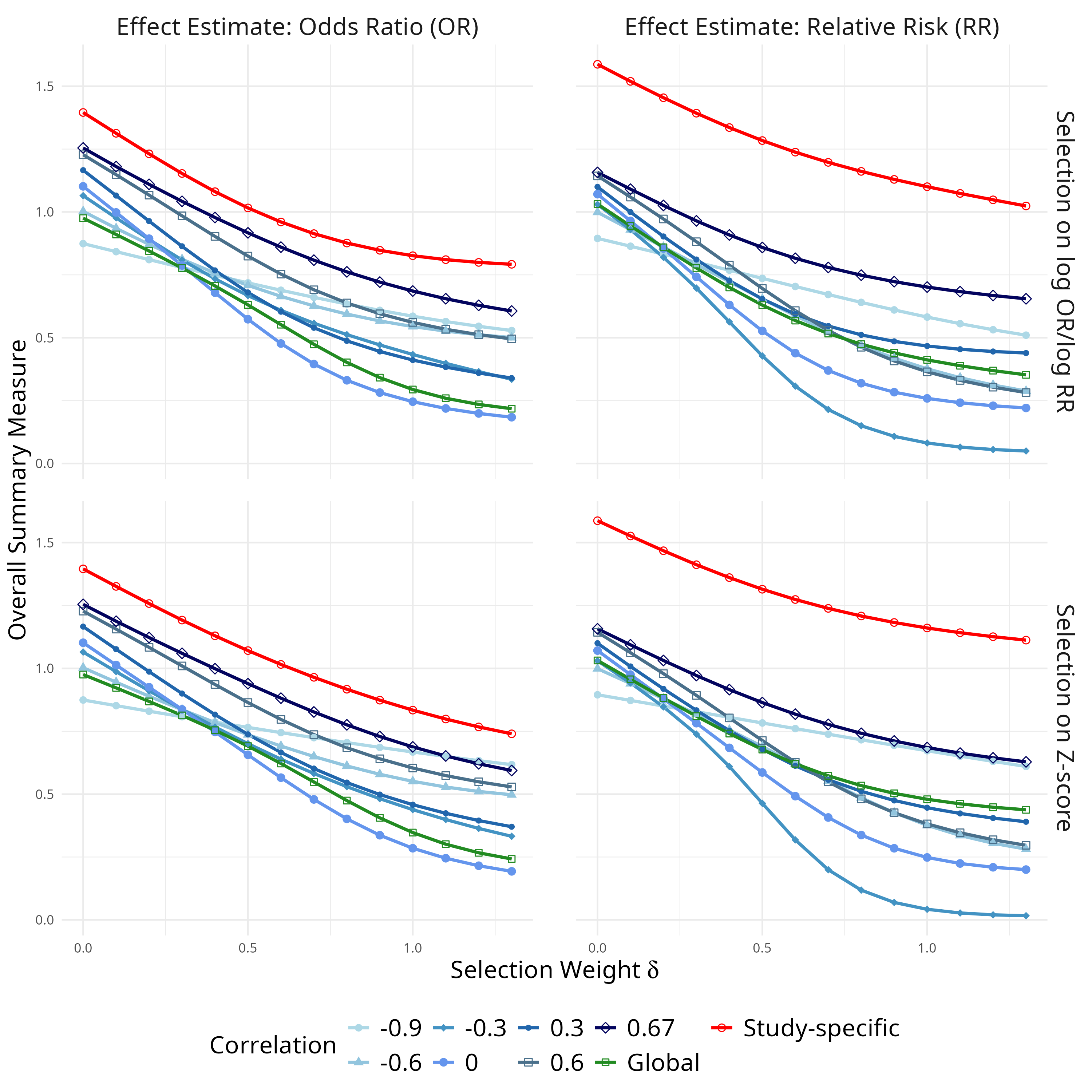}
\end{figure}

Figure \ref{fig:Multivariate1} compares the three approaches for incorporating within-study correlation in the multivariate imputation procedure. The assumed within-study correlation influences the imputed values and consequently the ORB-adjusted estimates. Differences between correlation assumptions are already visible at $\delta=0$, reflecting that the covariance structure affects the imputation even in the absence of selection. Under the study-specific correlation the borrowing generally becomes much less coherent and yields estimates that lie higher than those obtained under the fixed correlation, reflecting the additional uncertainty introduced by allowing correlations to vary across studies. As selection increases, the impact of the assumed correlation becomes more pronounced, particularly for \textit{seizure freedom} which has a larger proportion of unreported study outcomes. In contrast, results are more stable when few studies are unreported.

\begin{figure}[ht!]
\centering
\caption{Outcome \textit{seizure freedom}: comparison of univariate and multivariate ORB-adjustment (for a correlation of $r = -0.3$) for selection on the log OR/log RR and selection on the $z$-score over increasing selection.}
\label{fig:Uni_Multi}
\includegraphics[width = 1\linewidth]{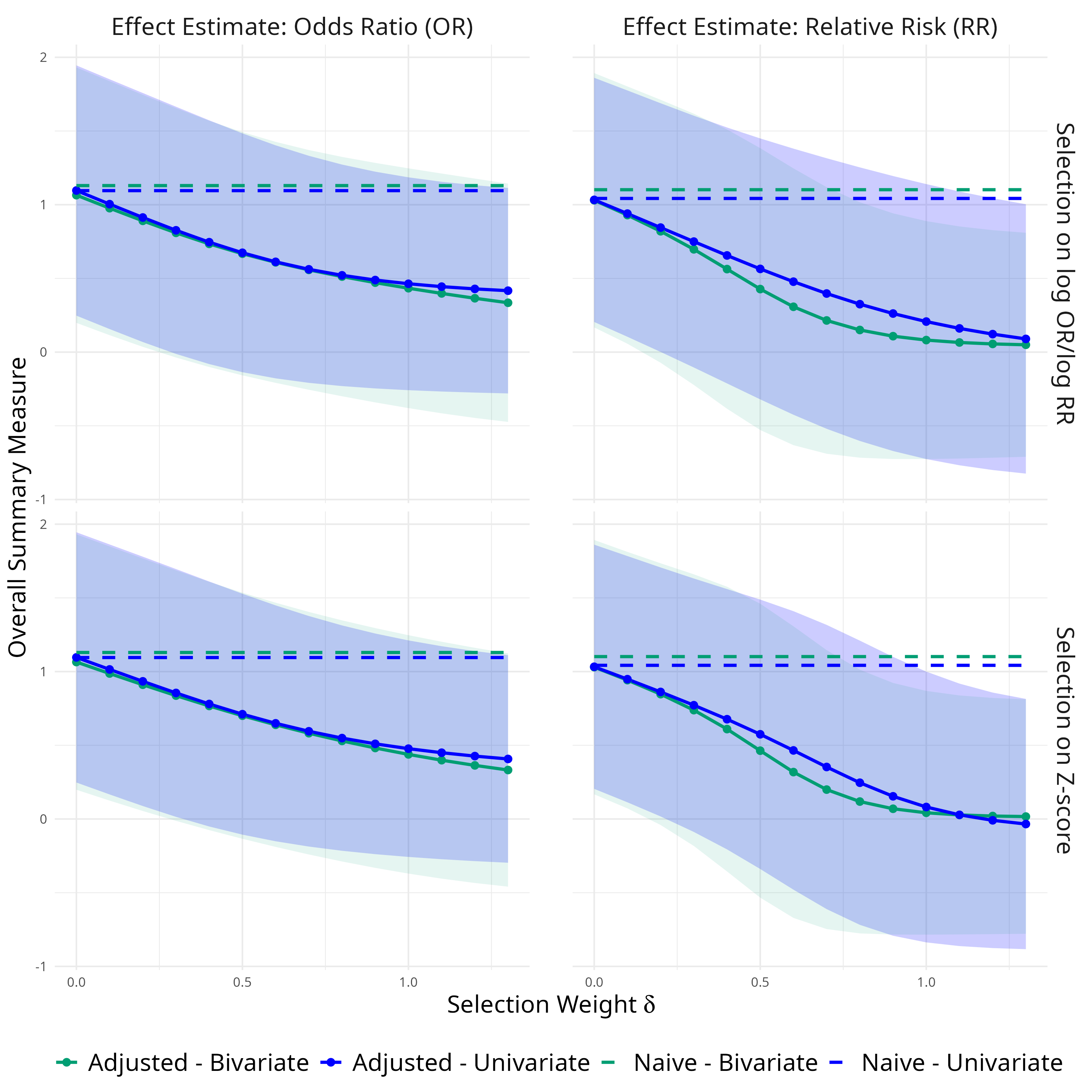}
\end{figure}

Figure \ref{fig:Uni_Multi} compares univariate and bivariate ORB-adjustments (fixed $r=-0.3$). The dashed lines, the naive estimates, refer to the MA estimates based only on the reported studies. These estimates differ between the univariate and bivariate approach because the multivariate model jointly incorporates both outcomes and their covariance structure. In all scenarios, the adjusted estimates decrease with increasing selection, reflecting stronger assumed selective outcome reporting. The bivariate adjustment was generally stronger than the univariate adjustment for larger selection weights, except for the log RR under $z$-score selection. We apply a single importance weight jointly across both outcomes within each imputation, consequently, favourable imputed treatment effects across outcomes reinforce each other through the shared selection mechanism. Imputations with jointly large treatment effects receive smaller weights, leading to stronger attenuation of the pooled estimates. The stronger bivariate adjustment illustrates how borrowing strength across correlated outcomes can amplify the impact of selective reporting when selection is modeled jointly across outcomes. In particular, the more completely reported \textit{50\% seizure reduction} outcome contributes additional information, which influences both the imputation model and the joint selection weights which creates a stronger adjustment compared to the outcome-specific univariate analysis. 

Although both \texttt{rma()} and \texttt{rma.mv()} can yield equivalent results under certain conditions, joint estimation changes the likelihood and weighting structure, leading to systematic differences even when $r=0$. These differences are more pronounced when many studies are unreported, highlighting the importance of jointly modeling outcome dependence and selection.

\section{Simulation Study}
\label{SimulationStudy}

We evaluate the ORB-adjustment under varying levels of heterogeneity and unreporting using a univariate and bivariate REM with selective reporting. Continuous outcomes are used for computational convenience, enabling straightforward interpretation of bias and coverage. Selection is modeled either on the $z$-score or on the treatment effect, and outcomes are used solely as statistical objects, so no clinical interpretation of the simulated outcomes is intended. The full simulation study protocol is available on \href{https://osf.io/5f4yx/overview}{OSF} and in the \href{https://github.com/cburgwin/ORB-project-MI-Approach/tree/main}{GitHub repository}.

\subsection{Setup}

Each simulated meta-analysis consists of $K \in \{6,12,25\}$ studies and two outcomes per study. Within-study sample sizes are fixed at $n_i=50$ per arm (total sample size per study $i$ given with $n_{ti} + n_{ci} =100$), consistent with typical MA simulation settings \cite{inthout2014hartung, saracini2025addressing}. The study-specific true treatment effects are generated from a bivariate normal distribution described in Section \ref{multivariate}: $\boldsymbol{\theta}_i \sim \Nor(\boldsymbol{\theta}, \boldsymbol{\Psi})$ where $\boldsymbol{\Psi}$ contains the between-study variances $\tau^2_1$, $\tau^2_2$ and between-study correlation $\rho_B$. The within-study covariance matrices are generated using the Wishart distribution $\boldsymbol{\Sigma}_i \sim \mathcal{W}_2(d, \boldsymbol{V}),$ where $\mathcal{W}_2$ denotes the 2-dimensional Wishart distribution \cite[p.~366]{held2020likelihood}, with degrees of freedom $d = 2(n_i - 1)$ and scale matrix
$\boldsymbol{V} = \frac{1}{(n_i - 1)n_i}
\begin{pmatrix}
1 & \rho_W \\
\rho_W & 1
\end{pmatrix}.$ Here, $\rho_W$ represents the within-study correlation between the two estimated effects. This formulation ensures that the marginal variances $\sigma_1^2$ and $\sigma_2^2$ are equal to $2/n_i$. Since we assume equal study sizes $n_i$ across all studies, $d$ and $\boldsymbol{V}$ remain constant for all $i$. Conditional on the true study-specific effects, the observed effect estimates are drawn from a bivariate normal distribution: $\widehat{\boldsymbol{\theta}}_i | \boldsymbol{\theta}_i \sim \Nor(\boldsymbol{\theta}_i, {\boldsymbol{\Sigma}}_i)$. Heterogeneity is parameterised via $I^2 \in \{0\%, 30\%, 60\%, 90\%\}$, corresponding to $\tau_j^2 \in \{0,0.02,0.06,0.36\}$. This allows scenarios ranging from no to high heterogeneity.

Selective reporting is imposed only on outcome 1 via a logistic selection model: $\Pr(R_{i1}=1 \mid s_{1}) = \operatorname{expit}(\alpha_1 + \delta_{1,\text{sim}} s_{1}),$ where $R_{i1}\in\{0,1\}$ is the reporting indicator, $\alpha_1$ controls the baseline reporting rate for outcome $1$ and $s_{1}$ is the selection variable either on the $z$-score or the treatment effect. The parameter $\delta_{1,\text{sim}}$ controls the selection strength, and $\delta_{1,\text{sim}}=0$ corresponds to a special MAR case, implying that selection has no influence on inference. In our setup, this also corresponds to missing completely at random (MCAR), since no covariates drive the reporting process, thus, the MAR and MCAR scenarios coincide. We assume one-sided selection favouring positive treatment effects.

To ensure a consistent degree of unreported study outcomes across different scenarios, $\alpha_1$ is calibrated to target a desired fraction $1 - p_1$ of reported studies for outcome $1$. Specifically, this is achieved by choosing $\alpha_1$ to satisfy
$\alpha_1 = \operatorname{logit}\!\left(1 - p_1 \right) - \delta_{1,\text{sim}} \mathbb{E}[s_{i1}]$. For selection on the $z$-score, ($s_{i1} = z_{i1}$), $\mathbb{E}[s_{i1}]$ denotes the expected standardized effect under the data-generating model
\begin{equation*}
    \mathbb{E}(z_{i1}) = 
    \frac{\mathbb{E}(\hat \theta_{i1})}{\sqrt{\mathbb{E}(\hat\sigma^2_{i1}) + \tau^2_1}} =
    \frac{\theta_{1}}{\sqrt{2/n_i + \tau^2_1}}.
\end{equation*}
For selection on the treatment effect ($s_{i1} = \hat\theta_{i1}$), $\mathbb{E}[s_{i1}] = \theta_1$. This ensures that, on average, the proportion of reported outcomes matches the target reporting rate $1 - p_1$, while allowing the individual reporting probability to vary with $s_{i1}$.
It is important to note that while our simulation uses a known missing data mechanism, \ie selective reporting based on a continuous function of the (standardized) observed treatment effect, no information about the cause of missingness is assumed to be available once the data are generated. This setup mimics the real-world challenge where outcomes are missing, but researchers lack the necessary ORBIT classification (\eg high risk/ low risk of bias) to inform the adjustment. Consequently, methods like Copas adjustment method \cite{copas2019model}, which require the ORBIT classification and adjust only for outcomes flagged as high risk of bias, are not applicable or directly comparable to our approach.

Meta-analyses were generated under different simulation scenarios (see Table \ref{tab:SimStudy} for an overview); \eg different true underlying treatment effects $\theta_j \in \{0, 0.4\}$ are considered, based on simulation studies found in the literature \cite{kirkham2012multivariate, saracini2025addressing}. We use a full factorial design, leading to 13824 scenarios. Each scenario is replicated $n_{\text{sim}}=1900$ times, and each replicate uses $M=200$ imputations (reduced from $M= 1000$ for computational feasibility as previous results were insensitive to this choice). Restricted maximum likelihood (REML) was used to estimate heterogeneity, as it is widely known and available in standard statistical software packages \cite{langan2019comparison, jackson2011multivariate, schmid2020handbook}. To assess model misspecification, we vary $\delta_{1,\text{sim}}$ and $\delta_{1,\text{est}}$ independently. Thereby, MAR (and MCAR) scenarios are included to evaluate the robustness of the ORB-adjustment method under alternative missingness mechanisms.

To avoid running into convergence issues and resulting missingness in the statistical analysis, additional control parameters have been specified in the \texttt{rma()} and \texttt{rma.mv()} functions of the \texttt{metafor} package \ie \texttt{control = list(stepadj = 0.1,  rel.tol=1e-8, maxiter = 1000)}.  In some simulation replicates, model fitting may still fail due to convergence issues. We repeat failed replicates until we obtain $n_{sim}$ successful replicates per scenario. This ensures that performance measures are computed on a fixed and comparable sample, reducing bias from arbitrary exclusions. All computations were performed using R version 4.5.1. The simulations were run in parallel using the \texttt{mclapply} function from the \texttt{parallel} R package on a 64-core Linux Server. 

We compare four estimates: (i) the naive estimate, (ii) the complete data estimate, (iii) the univariate ORB-adjusted estimate and (iv) the bivariate ORB-adjusted estimate. The naive estimate is obtained by fitting the bivariate REM to the subset of study outcomes that were reported. In practice, this corresponds to applying standard MA methods to the available data, ignoring the missing outcomes. The  complete  data  estimate  uses  all  studies  in the  meta-analysis  before  ORB  is  simulated which serves as a baseline for the true treatment effect in each scenario if there was no bias. It is also calculated using a bivariate REM and the ORB-adjusted estimates follow the theory explained in Section \ref{Importance}. Performance is evaluated using bias, mean squared error (MSE), coverage, and CI width.

\begin{table}[]
\captionsetup{justification=centering}
\caption{Simulation Setup Overview}
\label{tab:SimStudy}
\begin{adjustbox}{width=1\textwidth, center=\textwidth}
\begin{tabular}{|l|l|l|}
\hline
\textbf{Name}              & \textbf{Investigated values}  & \textbf{Description} \\ \hline
$K$    & 6, 12, 25    & Number of studies in the meta-analysis.  \\ \hline
$p_1$  & \begin{tabular}[c]{@{}l@{}} 20\%, 40\% \\\end{tabular} & \begin{tabular}[c]{@{}l@{}}Proportion of unreported studies for outcome 1.\end{tabular}      \\ \hline
$\tau_j^2$    & 0, 0.02, 0.06, 0.36     & \begin{tabular}[c]{@{}l@{}}Between-study heterogeneity variances for outcomes $j = 1, 2$.\end{tabular}   \\ \hline
$\theta_1$    & 0, 0.4    & \begin{tabular}[c]{@{}l@{}}True overall treatment effect for outcome 1.\end{tabular}    \\ \hline
$\theta_2$    & 0, 0.4    & \begin{tabular}[c]{@{}l@{}}True overall treatment effect for outcome 2.\end{tabular}    \\ \hline
$\rho_B, \rho_W$ & 0, 0.4 & Equal between-study and within-study correlation.
\\ \hline 
$\delta_{1,\text{sim}}$     & 0 - 1 in steps of 0.2    & \begin{tabular}[c]{@{}l@{}}Selection weight in simulation is varied to assess its\\ impact on the  adjusted estimates.\end{tabular}     \\ \hline
$\delta_{1,\text{est}}$     & 0 - 1 in steps of 0.2    & \begin{tabular}[c]{@{}l@{}}Selection weight in estimation is varied to assess its\\ impact on the adjusted estimates.\end{tabular}     \\ \hline
$s_{1}$     & $z_1$, $\hat\theta_1$    & \begin{tabular}[c]{@{}l@{}}Selection variable for either selection on\\ the $z$-score or selection on the treatment effect.\end{tabular}     \\ \hline
$M$    & 200  & Number of imputations for imputing the unreported studies.    \\ \hline
$n_{\text{sim}}$    & 1900  & Number of simulations for iterating over all scenarios.    \\ \hline
\end{tabular}
\end{adjustbox}
\end{table}

\subsubsection{Changes from the Simulation Study Protocol}
The simulation study followed the prespecified protocol with minor modifications mainly aimed at improving computational feasibility. First, the grid of correlation parameters was reduced, the range of selection weights was revised, and an additional selection mechanism based on the treatment effect was incorporated. Second, the number of imputations used within the multiple imputation procedure was reduced from 1000 to 200. Finally, the CI width was added as an additional performance metric and the univariate adjusted ORB estimate was included as an additional comparator. 

Preliminary simulation runs indicated that repeated estimation of heterogeneity parameters represented the primary computational burden, particularly for the bivariate models that were repeatedly fitted across multiple imputations. To reduce computation time, the variance components were treated as known and fixed at their generating values throughout the simulation study. Specifically, the univariate models were fitted using the true value of $\tau^2$, while the bivariate models were fitted using the true values of both $\tau_j^2$ and the between-study correlation parameter $\rho_B$. Consequently, only the treatment effect parameters were estimated during model fitting. Because the variance components were fixed, optimization was substantially simplified and less stringent convergence controls could be employed (\texttt{control = list(rel.tol=1e-5, maxiter = 200)}), resulting in a reduction in computational time. All other aspects of the data-generating mechanism and ORB-adjustment remained unchanged.

\subsection{Results}

We differentiate between two selection scenarios: selection based on the $z$-score and selection based on the treatment effect. We acknowledge that in real-world settings, selection on the $z$-score is often considered more plausible, as it directly reflects selection on the statistical significance, and studies are frequently withheld from publication due to a lack of significant results. However, by including both scenarios, we assess the performance of the ORB-adjustment method under different assumptions about the mechanism driving selective reporting. 

\subsubsection{Performance of Naive Estimate}
The naive estimate is substantially biased under ORB, with severity increasing under stronger selection and higher heterogeneity, \ie $I^2 = 90\%$ (see Figure \ref{fig:Bias}). The magnitude of the bias decreases as the true treatment effect size $\theta$ increases, which aligns with existing literature \cite{copas2019model, van2024correcting, saracini2025addressing}. Study size variations ($K = 6, 12, 25$) and the proportion of missingness, \eg whether 20\% or 40\% of studies are unreported, do not significantly affect the bias. For selection on the $z$-score, the bias is larger compared to selection on the treatment effect (see Figure \ref{fig:Bias} A) since highly positive but imprecise estimates (typically from smaller studies) are preferentially included while excluding equally imprecise, non-significant estimates. Regarding the CI width, the naive estimate generally exhibits wider CIs than the ORB-adjusted estimates because only the reported studies contribute to the estimation. High heterogeneity ($I^2 = 90\%$) amplifies the selection bias (since large effects are more likely reported) and increases uncertainty in $\hat\theta$, which leads to wider CIs. The coverage decreases for settings with higher heterogeneity. The type of selection has a substantial impact on the coverage, as the coverage decreases more for selection on the $z$-score (see Figure \ref{fig:Coverage} A). For high heterogeneity ($I^2$ = 90\%), the decrease in coverage gets stronger the more studies in the meta-analysis. 

\subsubsection{Bias with ORB-adjustment}

\begin{figure}[!ht]
\centering
\caption{Simulation results for bias of four MA estimates under ORB for $\theta_1 = \theta_2 = 0.4$, $\rho_B = \rho_W = 0.4$ and $p_i = 0.2$. The bias is shown for varying meta-analysis study sizes, heterogeneity levels, selection type and an increasing selection weight.} 
\label{fig:Bias}
\includegraphics[width = 1\linewidth]{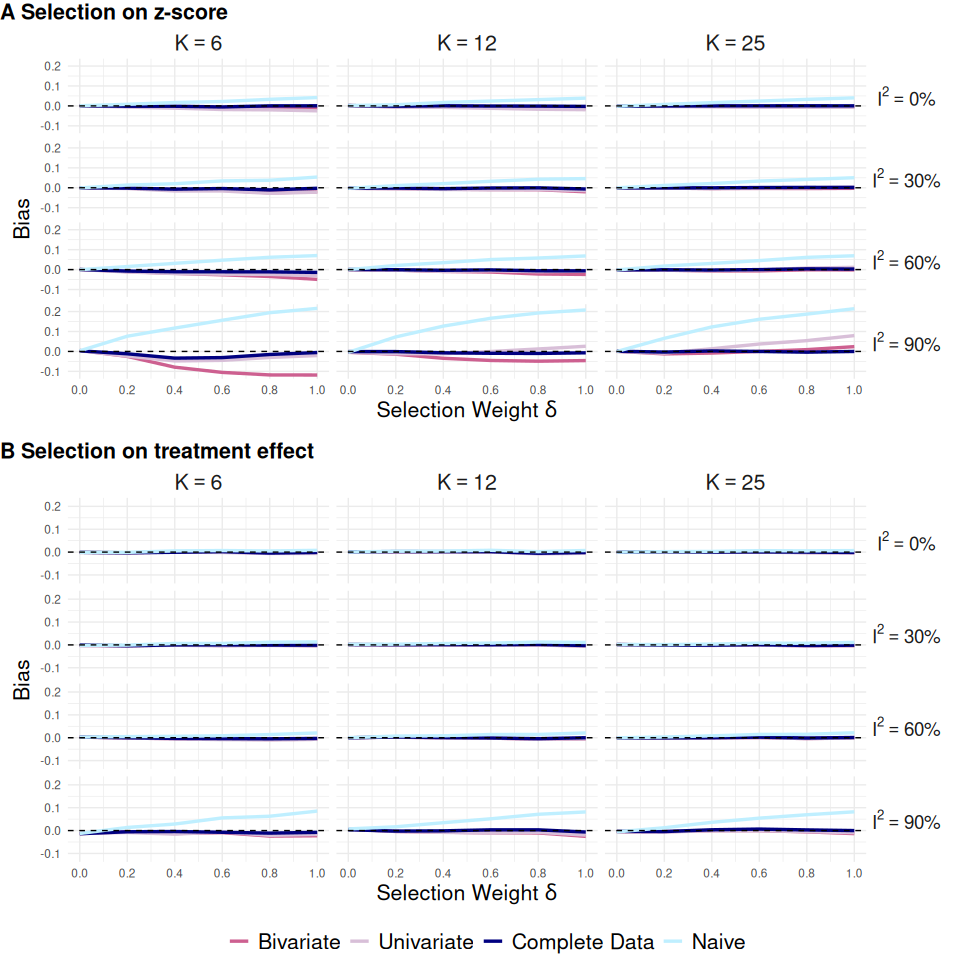}
\end{figure}

In comparison to the naive estimate, the bias reduces for the bivariate ORB-adjusted estimate by borrowing strength of the always reported outcome 2. The ORB-adjusted estimates shift the bias towards the null but for the high heterogeneity setting ($I^2 = 90\%$) fail to always eliminate it. The strength of correlation has a substantial impact, as for $\rho_B = \rho_W = 0$, the bias is higher compared to $\rho_B = \rho_W = 0.4$. Study size variations do not significantly affect the bias, the bias is just slightly larger for $K = 6$ for $I^2 = 90\%$ for the ORB-adjusted bivariate estimate. The bias for the univariate ORB-adjusted estimate is as large as for the naive estimate and is around 0.2 for $I^2 = 90\%$. Overall, the bivariate approach outperforms the univariate approach by borrowing strength across correlated outcomes, and the higher the correlation, the better the performance measures, \ie lower bias and higher coverage (see Figures \ref{fig:Bias} and \ref{fig:Coverage}). The bias for selection on the treatment effect is much smaller compared to selection on the $z$-score because it ignores study precision and does not preferentially filter out non-significant findings, thus creating a milder distortion of the effect distribution. Consequently, this leaves hardly any bias, even for high heterogeneity for both univariate and bivariate ORB-adjusted estimates (see Figure \ref{fig:Bias} B).
The slight negative bias observed for the ORB-adjusted estimates when $K = 6$ and high heterogeneity ($I^2 = 90\%$) suggests a small degree of overcorrection. With few studies available, both the imputation model and the importance sampling weights are estimated with greater uncertainty, making the adjustment more sensitive to random variation in the observed data. This effect decreases as the number of studies increases, leading to more stable weighting and estimates that are closer to the true treatment effect. The magnitude of the bias decreases as the true treatment effect size $\theta$ increases \cite{copas2019model, van2024correcting, saracini2025addressing}, \ie for $\theta_1 = \theta_2 = 0$ the bias is roughly twice as large as for $\theta_1 = \theta_2 = 0.4$. For unequal treatment effects, bias was largest for $\theta_1 = 0$ and $\theta_2 = 0.4$, likely because ORB operated on the first outcome, whose true effect was null. The bivariate model borrows strength from outcome 2, but outcome 2 is centered at a different mean. Consequently, the information borrowed from the second outcome is less representative of the first outcome than in scenarios where both outcomes have the same treatment effects, leading to greater bias. The proportion of unreporting has hardly any influence on the ORB-adjustment for $\delta_{1,\text{est}} = \delta_{1,\text{sim}} \neq 0$. 

\subsubsection{Other Performance Measures for ORB-adjusted Estimate}

\begin{figure}[!ht]
\centering
\caption{Simulation results for the confidence interval width of four MA estimates under ORB for $\theta_1 = \theta_2 = 0.4$, $\rho_B = \rho_W = 0.4$ and $p_i = 0.2$. The confidence interval width is shown for varying meta-analysis study sizes, heterogeneity levels, selection type and an increasing selection weight.}
\label{fig:CI_Width}
\includegraphics[width = 1\linewidth]{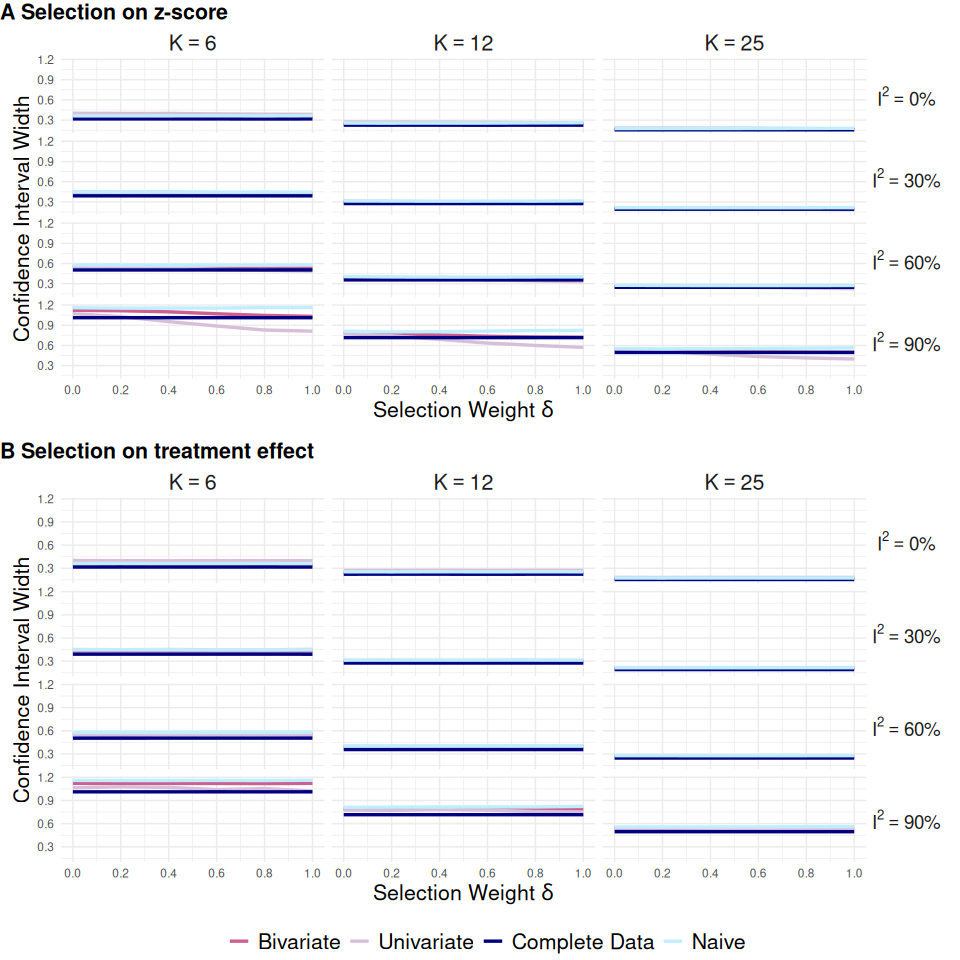}
\end{figure}

\begin{figure}[!ht]
\centering
\caption{Simulation results for the MSE of four MA estimates under ORB for $\theta_1 = \theta_2 = 0.4$, $\rho_B = \rho_W = 0.4$ and $p_i = 0.2$. The MSE is shown for varying meta-analysis study sizes, heterogeneity levels, selection type and an increasing selection weight.} 
\label{fig:MSE}
\includegraphics[width = 1\linewidth]{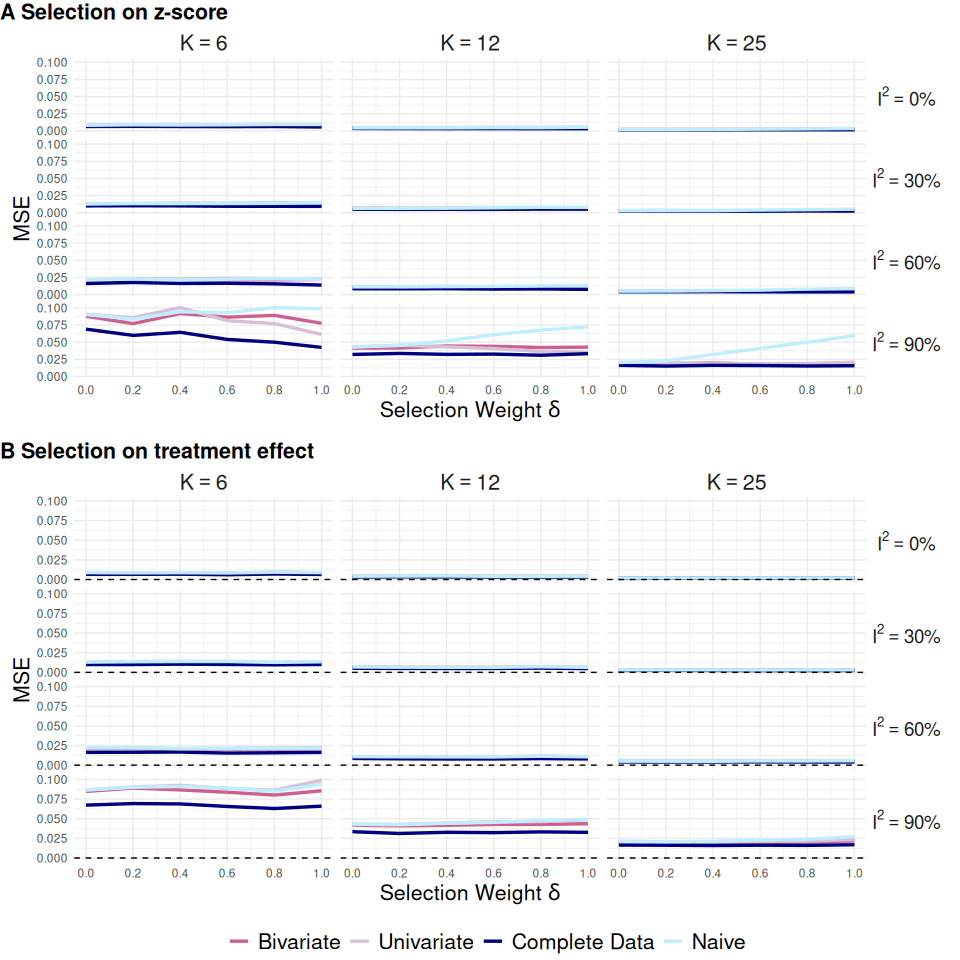}
\end{figure}

For all considered scenarios, the CI width decreases as the number of studies, $K$, increases because more studies contribute a larger total weight to the meta-analysis, leading to a smaller standard error for the pooled effect estimate. This effect is especially pronounced in the high heterogeneity setting ($I^2 = 90\%$) because the larger $K$ allows for a more stable and accurate estimation of the high between-study variance. The complete data estimate yields the lowest CI width (see Figure \ref{fig:CI_Width}), indicating higher precision compared to the naive estimate and the ORB-adjusted estimates, except for selection on the $z$-score and high heterogeneity. The CI width is observed to be slightly smaller for larger true treatment effects. Crucially, the differences in CI width between the estimate types are stronger and more pronounced for high heterogeneity ($I^2 = 90\%)$.

The MSE also decreases as the number of studies increases, as the larger study size improves the precision. For selection on the $z$-score and high heterogeneity, the MSE for the naive estimate increases over increased selection (see Figure \ref{fig:MSE} A). The complete data estimate consistently has the lowest MSE. The ORB-adjusted estimates show visible numerical instability when the number of studies is small ($K = 6$) and selection is based on the $z$-score. In this setting, high sampling variance, strong selection effects under substantial heterogeneity, and variability in the importance sampling weights together produce less stable imputations and greater dispersion of the adjusted estimates, leading to higher MSE despite the reduction in bias.

\begin{figure}[!ht]
\centering
\caption{Simulation results for the coverage of four MA estimates under ORB for $\theta_1 = \theta_2 = 0.4$, $\rho_B = \rho_W = 0.4$ and $p_i = 0.2$. The coverage is shown for varying meta-analysis study sizes, heterogeneity levels, selection type and an increasing selection weight.}
\label{fig:Coverage}
\includegraphics[width = 1\linewidth]{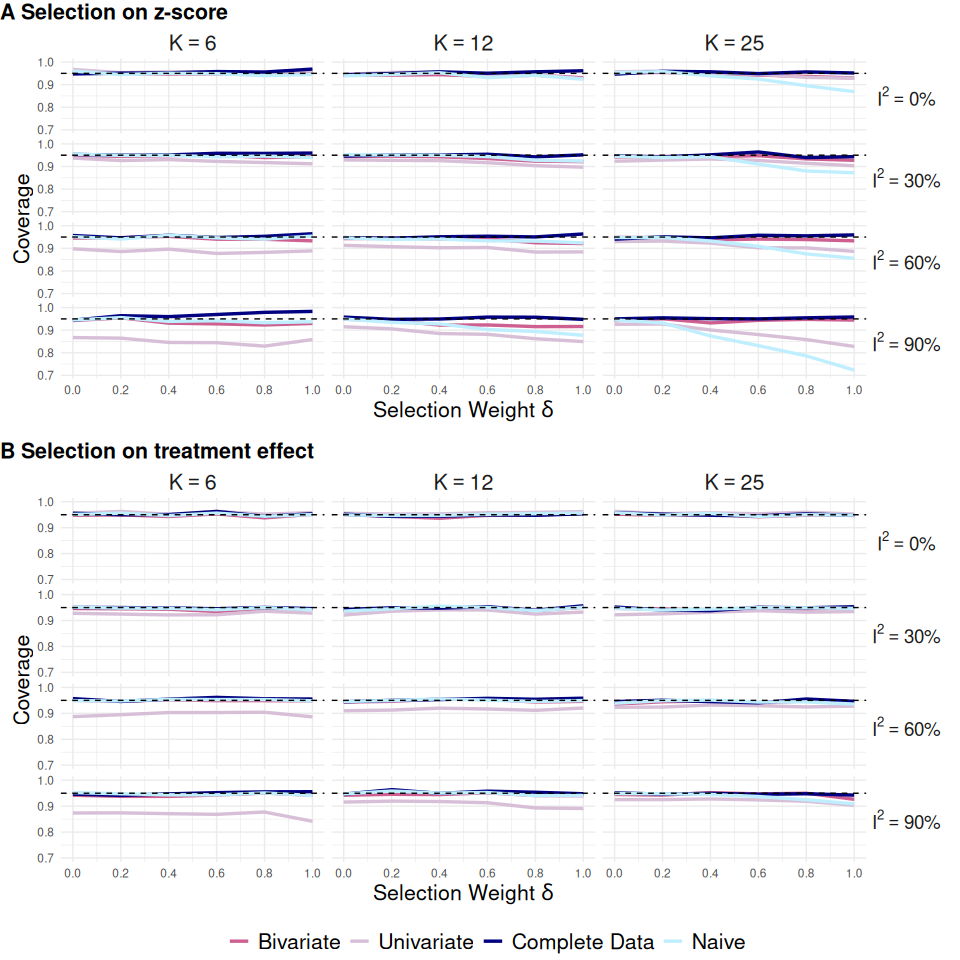}
\end{figure}

The coverage decreases for the univariate ORB-adjusted estimate for increasing selection weight for the high heterogeneous setting (see Figure \ref{fig:Coverage}). The coverage is lowest for the naive estimate for selection on the z-score (see Figure \ref{fig:Coverage} A). The coverage is lower for smaller true treatment effects for selection on the $z$-score. Conversely, coverage is lower for unequal treatment effects, \ie the coverage is lower for $\theta_1 = 0$ and $\theta_2 = 0.4$ compared to $\theta_1 = \theta_2 = 0.4$. Although the outcomes were moderately correlated ($\rho_B = \rho_W = 0.4)$ the outcome affected by ORB had a null mean, while the second outcome had a positive mean. This reduced the representativeness of the borrowed information relative to scenarios in which both outcomes had similar means, resulting in greater bias and hence lower coverage.
For $\rho_B = \rho_W = 0$, the coverage drops significantly more for the bivariate ORB-adjusted estimate compared to $\rho_B = \rho_W = 0.4$. For $\rho_B = \rho_W = 0$ the coverage of the bivariate ORB-adjusted estimate is worse than the coverage of the univariate ORB-adjusted estimate. The lower coverage of the bivariate approach when $\rho_B = \rho_W = 0$ is expected because the second outcome provides no information about the first outcome and as a result, borrowing strength is ineffective. 

\subsubsection{Model Misspecification}
We also evaluated the robustness of our ORB-adjustment method under alternative missingness mechanisms by varying $\delta_{1,\text{sim}}$ and $\delta_{1,\text{est}}$ independently. 
In scenarios where no selection in the data generation occurred ($\delta_{1,\text{sim}} = 0$) but an adjustment was applied ($\delta_{1,\text{est}} \neq 0$), the estimates remain largely unbiased. As selection in the data generation increased ($\delta_{1,\text{sim}} \geq 0.6$) and $\delta_{1,\text{est}} \neq 0$, the bias is slightly stronger for $K = 6$ and $I^2 = 90\%$. The coverage for both the naive and univariate ORB-adjusted estimate drops over increased heterogeneity. For the bivariate ORB-adjusted estimate, again a stronger correlation between the outcomes and therefore, stronger borrowing of strength from the fully reported outcome 2 results in lower bias and no decrease in coverage for the bivariate ORB-adjusted estimate.
For selection on the treatment effect, the bias was less severe. For example, for $\delta_{1,\text{sim}} = 0.2$ and increasing $\delta_{1,\text{est}}$, the selection weight is misspecified in the analysis, but the true selection mechanism is only weakly MNAR which results in a weak bias. Overall, however, for both selection types, the bias is stronger, and the coverage lower the further the selection weights are apart from each other (see supplementary material for more results).

\subsubsection{Numerical stability}
To evaluate numerical stability across the simulation grid, we monitored the model non-convergence rate for the bivariate multiple imputation and the total number of attempts required to reach $1,900$ successful replications. Across all evaluated scenarios, the target simulation size was secured without approaching the maximum safety ceiling of $9,500$ attempts ($5 \times n_{\text{sim}}$). The diagnostic metrics revealed that numerical stability is  governed by the number of studies in the meta-analysis. For example for the scenario considered in Figures \ref{fig:Bias} - \ref{fig:Coverage}: the largest considered sample size ($K = 25$) demonstrated almost perfect execution, with failure rates below 1\%. In contrast, for $K = 6$ the data sparsity introduced by ORB required an average of 0.13 to 0.18 data regeneration redraws per successful loop to ensure the minimum threshold of 4 reported studies. Furthermore, optimizing these small sample imputed datasets resulted in a  bivariate model non-convergence rate of 26\% to 33\% across all levels of heterogeneity, requiring between 2,580 and 2,850 total attempts to achieve the target simulation size of 1900. To conclude,  small MA sample size, rather than heterogeneity, represents the primary constraint for numerical convergence for the bivariate adjustment approach.

\section{Discussion}
\label{discussion}

There has been little consideration how to adjust for ORB in meta-analyses. Because the reporting mechanism is generally unknown and depends on unobserved study results, ORB constitutes a non-ignorable missing data problem. We propose an ORB-adjustment method based on multiple imputation and use selection models that allow both effect estimate and $z$-score based selection. Extending previous work by Carpenter et al. \cite{carpenter2011assessing} and Williamson and Gamble \cite{williamson2005identification}, the method can be applied within both univariate and multivariate meta-analysis frameworks, allowing borrowing of strength across correlated outcomes.

We applied the proposed ORB-adjustment methodology to a real-world meta-analysis of epilepsy trials \cite{pulman2014topiramate, bresnahan2019topiramate} affected by ORB. The ORB-adjusted estimates of the treatment effect differed substantially compared to the naive estimates. This adjustment was particularly significant when half of the studies did not report any data for the outcome. This aligns with previous findings demonstrating that meta-analyses with a higher proportion of unreported or unpublished data exhibit substantially larger shifts in effect estimates, ultimately leading to greater over- or underestimation of treatment efficacy if left unadjusted \cite{saracini2025addressing, hart2012effect}. Selection on the $z$-score generally resulted in stronger adjustment than selection on the effect estimate, particularly under stronger assumed selection which is consistent with the findings from Carpenter et al. \cite{carpenter2007sensitivity}. In the multivariate setting, adjustment differed from the univariate case because correlated outcomes are incorporated both in the imputation model and in the importance sampling weights. Previous literature has shown that, in conventional multivariate meta-analysis, borrowing strength across outcomes can improve estimation efficiency and reduce standard errors by incorporating additional information across outcomes \cite{jackson2011multivariate, riley2007evaluation}. Our setting differs because the multivariate model is embedded within a multiple imputation approach followed by selection model reweighting. Consequently, the assumed correlation influences not only the imputed outcomes but also the selection weights assigned to each imputation. While borrowing strength can improve the information available for imputing unreported outcomes, the additional variability introduced through the weighting step may offset these efficiency gains. As a result, the multivariate ORB adjustment does not necessarily yield smaller standard errors than separate univariate analyses, particularly under strong selection or high heterogeneity.

The findings of our simulation study extend the results of the clinical data by investigating the impact of several parameter variations on ORB-adjustment. The simulation study investigated the performance of the proposed method across a range of scenarios, including model misspecification. Naive estimates display substantial bias, particularly in high heterogeneity settings, underscoring the importance of adjusting for ORB. Across the simulated scenarios, the proposed ORB-adjusted methods generally reduced average bias and MSE relative to the naive estimate and achieved coverage closer to the nominal level, although performance depended on the degree of heterogeneity, the strength of correlation, and the assumed selection mechanism. Selection based on the $z$-score  consistently resulted in more pronounced bias and subsequent adjustment compared to selection on the treatment effect. This is a direct consequence of the dependence of $z$-score selection on both the effect estimate and its standard error, which preferentially includes potentially exaggerated effects from small, imprecise studies. Previous simulation studies caution that simple weighting after multiple imputation in the setting of single variable missing data may not adequately correct for MNAR mechanisms, as bias can persist even with large samples and many imputations \cite{hayati2015evaluation, carpenter2007sensitivity}. In contrast, our proposed method embeds the weighting within a selection model for ORB, where the probability of reporting is parameterized as a function of study-level significance. Nevertheless, under strong heterogeneity ($I^2=90\%$) and a small number of studies, residual bias remained even after ORB adjustment.

Several limitations should be acknowledged. First, our ORB-adjustment method requires reported study sample sizes and the selection weight $\delta$ is generally not identifiable from the data, necessitating a sensitivity analysis approach. Second, the reporting mechanism was assumed to only depend on study results, whereas reporting decisions may also be influenced by study-level characteristics such as funding or publication practices. Third, in the simulation study the between-study variance and correlation parameters were treated as known and fixed at their generating values. In practice, these quantities are unknown and must be estimated from the observed data. Consequently, the simulation does not fully capture the uncertainty associated with estimating heterogeneity parameters, particularly in meta-analyses with a small number of studies. The reported performance measures may therefore be somewhat optimistic relative to real-world applications. Finally, the multivariate approach is computationally heavy, particularly when using iterative estimation methods like REML. 

Future research can investigate applying the ORB-adjustment method to harmful outcomes. Besides, for the multivariate approach computationally efficient methods can be explored to handle a larger number of outcomes. This will make our ORB-adjustment approach more appealing for network meta-analysis \cite{mavridis2014selection}. Network meta-analysis improves the traditional pairwise meta-analysis by combining multiple sources of evidence from a network of studies \cite{mills2013demystifying}. One could also investigate the behaviour of the proposed ORB-adjustment method for individual meta-analyses, beyond the average performance assessed in the present simulation study.
We further have only focused on multiple outcomes, but multivariate meta-analysis can also be applied to multiple treatment groups \cite{hasselblad1998meta}. 
Although complete IPD and analysis-ready datasets would largely eliminate ORB by allowing treatment effects to be estimated directly from the data, IPD would nevertheless be valuable for the proposed multivariate approach because it enables direct estimation of the within-study covariance \cite{jackson2011multivariate, kirkham2012multivariate, mavridis2013practical}. Furthermore, in practice IPD are often available only for a subset of studies, in which case selective availability of outcomes may still occur. Hybrid approaches combining IPD and aggregate data therefore represent an interesting direction for future research.
Another promising avenue for future research is exploring how to integrate the risk of bias classification from the ORBIT classification system \cite{kirkham2010impact}. For example, Copas et al. \cite{copas2014model} based their model-based adjustment method on the ORBIT classification. The risk of bias ratings could be used to inform the imputation models, for example, unreported study outcomes flagged as high risk could be imputed differently than those flagged as low risk. 

\section{Conclusions}
\label{conclusion}
This work demonstrates that ORB can substantially distort treatment effect estimates in meta-analysis, particularly when a large proportion of study outcomes is unreported. To address this, we propose a flexible ORB-adjustment approach based on multiple imputation and selection models that can be implemented in both univariate and multivariate meta-analysis. The multivariate extension allows borrowing of strength across correlated outcomes and can improve imputation, although the extent of borrowing depends on the assumed within-study correlation and the proportion of unreported study outcomes.

Application to a real-world epilepsy meta-analysis showed that ORB-adjustment shifted treatment effects towards the null compared with naive analyses, while the simulation study demonstrated reductions in bias and improved coverage across a broad range of scenarios, including moderate misspecification of the assumed selection mechanism. Overall, the proposed approach provides a practical sensitivity analysis for assessing the robustness of MA conclusions to ORB.

\subsection*{List of abbreviations}

\begin{table}[H]
\begin{tabular}{ll}
CI    & Confidence interval              \\
IPD   & Individual participant data      \\
MAR   & Missing at random                \\
MCAR  & Missing completely at random     \\
MA    & Meta-analytic                    \\
MNAR  & Missing not at random            \\
MSE   & Mean squared error               \\
OR    & Odds ratio                       \\
ORB   & Outcome reporting bias           \\
ORBIT & Outcome reporting bias in trials \\
OSF   & Open science framework           \\
RCT   & Randomized controlled trial      \\
REM   & Random-effects model             \\
REML & Restricted maximum likelihood     \\
RR    & Relative risk                    \\
SE    & Standard error                  
\end{tabular}
\end{table}

\section*{Declarations}

\subsection*{Ethics approval and consent to participate}
Not applicable.

\subsection*{Consent for publication}
Not applicable.

\subsection*{Availability of data and materials}
The datasets and code supporting the conclusions of this article are available in the \href{https://github.com/cburgwin/ORB-project-MI-Approach/tree/main}{ORB-project-MI-Approach GitHub} repository.

\subsection*{Competing interests}
The authors declare that they have no competing interests.

\subsection*{Funding}
CB and LH are members of the SHARE-CTD doctoral network on clinical trial data sharing (Horizon-MSCA.2022-DN 110120360), funded by the European Union. The authors are supported by the Swiss State Secretariat for Education, Research and Innovation (SERI) under subsidy contract No. 23.00303.

\subsection*{Authors’ contributions}
\noindent
CB: Conceptualization, Formal analysis, Methodology, Software, Visualization, Writing – original draft, Writing – review \& editing. SF: Methodology, Formal analysis, Software, Validation, Writing – review \& editing. LH: Conceptualization, Methodology, Writing – review \& editing, Supervision, Project administration, Funding acquisition. All authors read and approved the final manuscript.

\subsection*{Acknowledgements}
Not applicable.

\sloppy
\printbibliography

\newpage

\input{Appendix.tex}

\end{spacing}
\end{document}

%% file: newCommands.tex
\usepackage{xifthen}            % for \isempty
\usepackage{array}
\usepackage{amssymb}

\newcommand{\blanco}[1]{  } 
\newcommand{\Rep}{\mbox{\tiny{Rep}}}

\newcommand{\SE}{\mbox{{SE}}}
\newcommand{\Var}{\mbox{{Var}}}
\newcommand{\Adj}{\tiny{\mbox{Adj}}}
\newcommand{\MA}{\tiny{\mbox{MA}}}

\newcommand{\latin}[1]{\textit{#1}}

\newcommand{\abk}[1]{\mbox{#1}\xdot}
\DeclareRobustCommand\xdot{\futurelet\token\Xdot}
\def\Xdot{%
  \ifx\token\bgroup.%
  \else\ifx\token\egroup.%
  \else\ifx\token\/.%
  \else\ifx\token\ .%
  \else\ifx\token!.%
  \else\ifx\token,.%
  \else\ifx\token:.%
  \else\ifx\token;.%
  \else\ifx\token?.%
  \else\ifx\token/.%
  \else\ifx\token'.%
  \else\ifx\token).%
  \else\ifx\token-.%
  \else\ifx\token+.%
  \else\ifx\token~.%
  \else\ifx\token.%
  \else.\ %
  \fi\fi\fi\fi\fi\fi\fi\fi\fi\fi\fi\fi\fi\fi\fi\fi%
}

\newcommand{\eg}{\abk{\latin{e.\,g}}}
\newcommand{\ie}{\abk{\latin{i.\,e}}}

\newlength{\halbebreite}
\DeclareMathOperator{\Nor}{N} % Normal -
\DeclareMathOperator{\Cov}{Cov} % Covariance
\DeclareMathOperator{\diag}{diag} % Diagonalmatrix
\DeclareMathOperator{\arctanh}{arctanh} % arcus tangens hyperbolicus
\newcommand{\partialv}[3][1]{%
\ifthenelse{#1 = 1}{\frac{\partial #2}{\partial #3}}{\frac{\partial^{#1} #2}{\partial #3^{#1}}}
} 

\newcommand{\partials}[3][1]{%
\ifthenelse{#1 = 1}{\frac{d #2}{d #3}}{\frac{d^{#1} #2}{d #3^{#1}}}
} 

\newcommand{\dseps}[2][1]{%
\ifthenelse{#1 = 1}{\frac{d}{d #2}}{\frac{d^{#1}}{d #2^{#1}}}
}

\newcommand{\dsepv}[2][1]{%
\ifthenelse{#1 = 1}{\frac{\partial}{\partial #2}}{\frac{\partial^{#1}}{\partial #2^{#1}}}
}

\newcommand{\ml}[2][1]{% % für Maximum-Likelihood-Schätzer von #1
\ifthenelse{#1 = 1}%
 {\hat{#2}_{\scriptscriptstyle{\text{ML}}}}% 
 {\hat{#2}^{#1}_{\scriptscriptstyle{\text{ML}}}}% z.B. für sigmadach^2
}
\newcommand{\map}[2][0]{% % für MAP-Schätzer von #1
\ifthenelse{#1 = 0}%
 {\hat{#2}_{\scriptscriptstyle{\text{MAP}}}}% 
 {\hat{#2}_{{\scriptscriptstyle{\text{MAP}}_{#1}}}}% z.B. für sigmadach^2
}
\newcommand{\mpm}[2][0]{% % für MAP-Schätzer von #1
\ifthenelse{#1 = 0}%
 {\hat{#2}_{\scriptscriptstyle{\text{MPM}}}}% 
 {\hat{#2}_{{\scriptscriptstyle{\text{MPM}}_{#1}}}}% z.B. für sigmadach^2
}

\newcommand{\given}{\,\vert\,} % für "X gegeben Y" also $X\given Y$ schreiben

%% file: Appendix.tex
\section*{Supplementary Material}
\label{supp_results}

\maketitle

\subsection*{Application}

In the manuscript, we focused primarily on the outcome \textit{seizure freedom} as it has more unreported study outcomes and therefore provides a clearer visualization of our ORB adjustment method. For completeness, this section presents the corresponding plots for the second outcome \textit{50\% seizure reduction}.

\begin{figure}[ht!]
\centering
\caption{Outcome \textit{50\% seizure reduction}: comparison of naive and univariate ORB-adjusted estimates for selection on the log OR/ log RR and selection on the z-score over increasing selection.}
\label{fig:Univariate2}
\includegraphics[width = 1\linewidth]{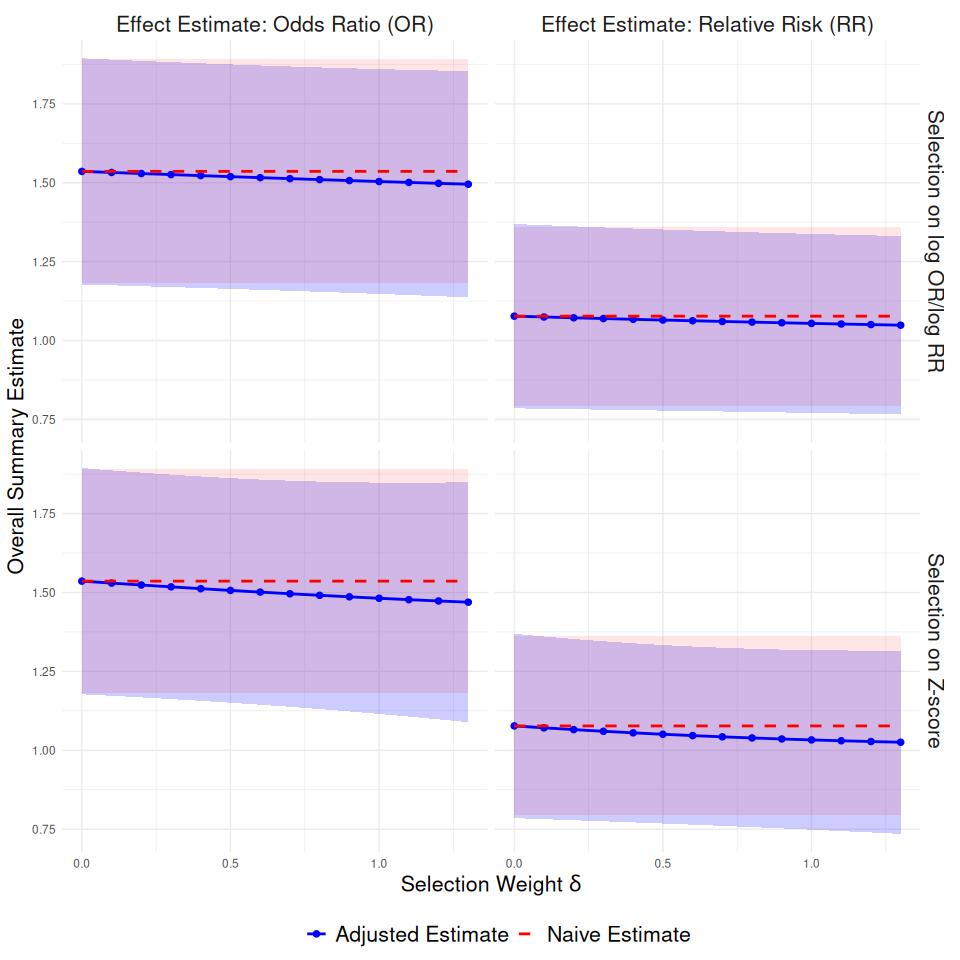}
\end{figure}

As expected, the ORB-adjusted treatment effects and corresponding CIs are generally shifted toward the null compared to the naive MA estimates based only on the reported studies (Figure \ref{fig:Univariate2}). However, because only one study outcome is unreported for \textit{50\% seizure reduction}, the magnitude of the adjustment is substantially smaller than for \textit{seizure freedom}, where approximately half of the study outcomes are unreported. Consequently, the CIs of the naive and adjusted estimates are very similar. For both the OR and RR, selection on the z-score produces slightly stronger adjustment than selection on the effect estimate. This reflects the dependence of z-score selection on both the treatment effect and its standard error, giving greater influence to more precise studies as the selection weight increases. Similar to the findings for \textit{seizure freedom}, the OR estimates remain larger and have wider CIs  than the RR estimates. Overall, however, the differences between naive and adjusted estimates remain modest because the proportion of unreported study outcomes is small.

\begin{figure}[ht!]
\centering
\caption{Outcome \textit{50\% seizure reduction}: comparison of univariate and multivariate ORB-adjustment (for a correlation of $r = -0.3$) for selection on the log OR/ log RR and selection on the z-score over increasing selection.}
\label{fig:Uni_Multi_O1}
\includegraphics[width = 1\linewidth]{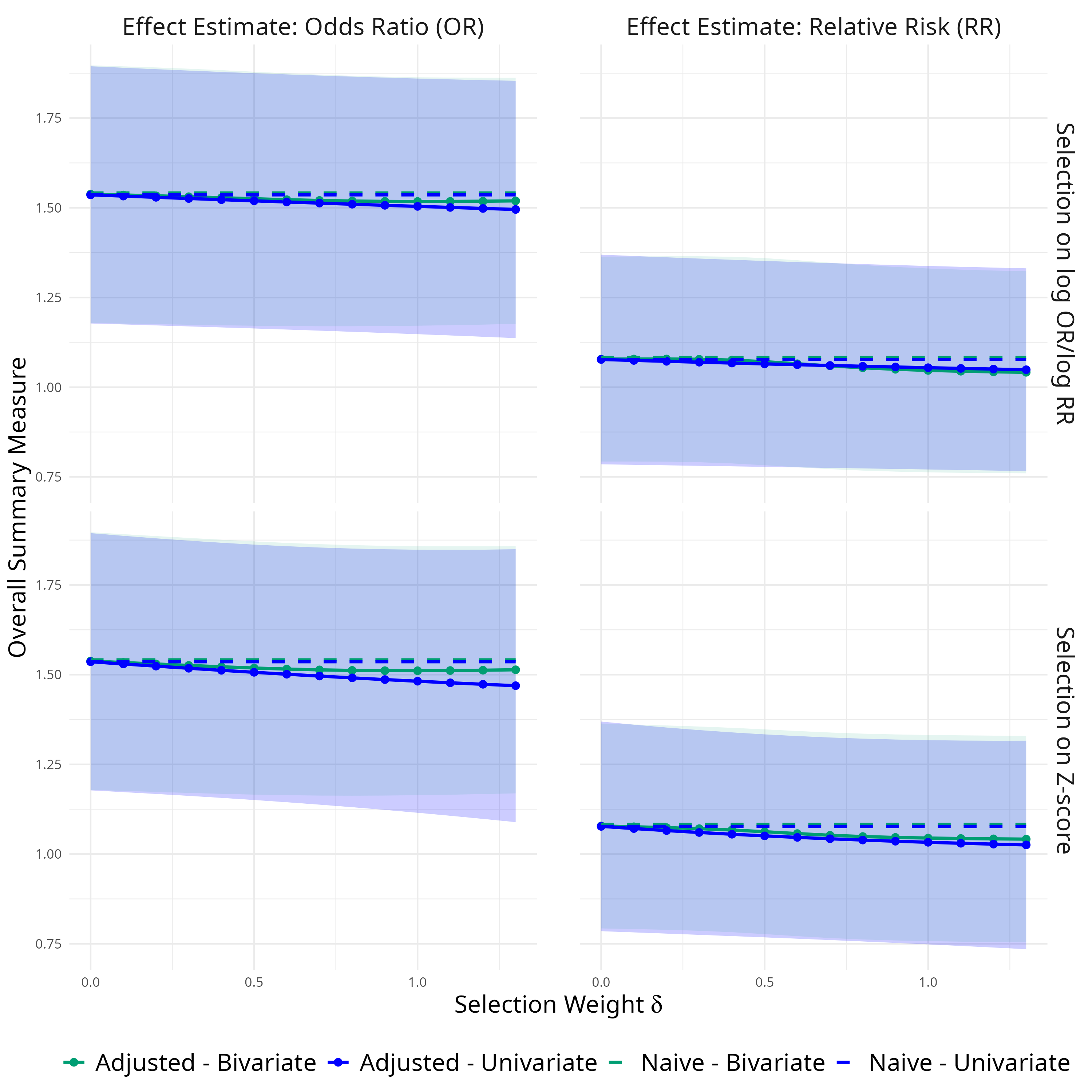}
\end{figure}

Next, we compared the univariate and multivariate ORB adjustment for the outcome \textit{50\% seizure reduction}, assuming a fixed within-study correlation of -0.3 (see Figure~\ref{fig:Uni_Multi_O1}). The dashed lines represent the naive MA estimates based only on the reported studies. These estimates differ between the univariate and multivariate approach because the multivariate model jointly incorporates both outcomes and their covariance structure. Across all scenarios, the adjusted estimates decrease with increasing selection, reflecting stronger assumed ORB. In contrast to the stronger differences observed for \textit{seizure freedom}, the univariate and multivariate adjustments for \textit{50\% seizure reduction} remain relatively similar, with largely overlapping CIs. For selection on the z-score, the univariate adjustment is slightly stronger than for selection on the effect estimate.

\begin{figure}[ht!]
\centering
\caption{Outcome \textit{50\% seizure reduction}: comparison of within-study correlations for multivariate ORB-adjustment for selection on the log OR/ log RR and selection on the z-score over increasing selection.}
\label{fig:Multivariate2}
\includegraphics[width = 1\linewidth]{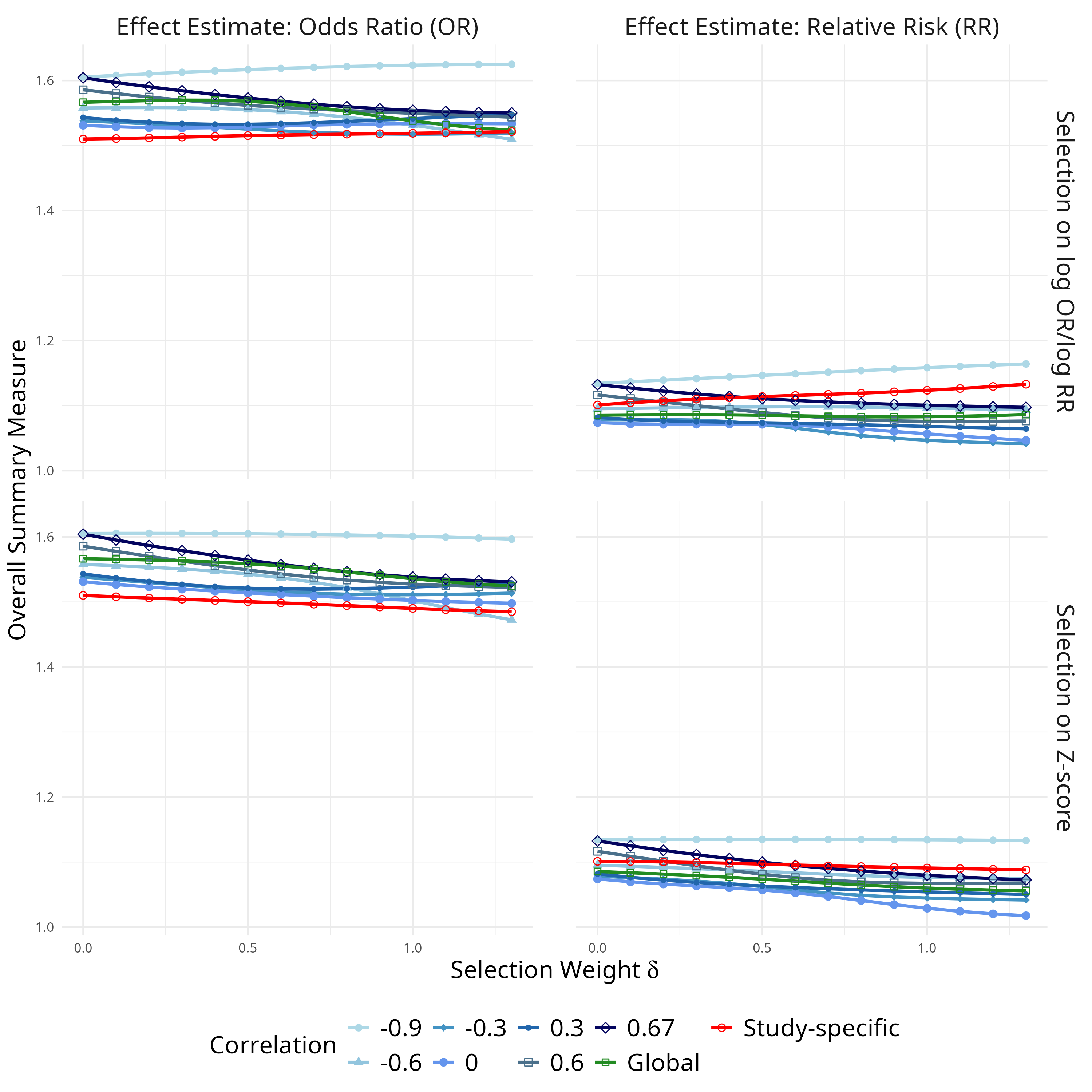}
\end{figure}

Figure \ref{fig:Multivariate2} compares the three approaches for incorporating within-study correlation for the outcome \textit{50\% seizure reduction}. In contrast to seizure freedom, only a single study outcome is unreported for this outcome. Consequently, the adjusted estimates are only weakly affected by the assumed within-study correlation, and the curves remain closely clustered across the full range of selection weights. Differences between correlation assumptions are already present at $\delta = 0$, reflecting the influence of the covariance structure on the imputed value under MAR. As the selection weight increases, the adjusted estimates change only modestly because the overall analysis is driven primarily by the reported outcomes. The study-specific correlation reflects the additional variability introduced by sampling a separate correlation for each study, although the overall differences remain small.

Overall, the results for \textit{50\% seizure reduction} demonstrate that the magnitude and stability of ORB adjustment depends strongly on the proportion of unreported study outcomes. When only a small proportion of outcomes is unreported, both univariate and multivariate adjustments remain relatively modest and robust to different modeling assumptions. In contrast, outcomes with a larger proportion of missing information are substantially more sensitive to assumptions regarding the selection mechanism and the within-study correlation structure. These findings highlight that multivariate ORB adjustment is most influential in settings with substantial selective non-reporting, where borrowing strength across correlated outcomes can alter the adjusted estimates and their uncertainty.

\subsection*{Simulation Study}
In the manuscript’s results section of the simulation study, we focus primarily on scenario where $\delta_{1,\text{sim}} = \delta_{1,\text{est}}$. This supplementary document presents additional results from the simulation study focusing on $\delta_{1,\text{sim}} \neq \delta_{1,\text{est}}$. 

\begin{figure}[!ht]
\centering
\caption{Comparison of estimates for $\theta_1 = \theta_2 = 0.4$, $\rho_B = \rho_W = 0.4$, $p_i = 0.2$ and $\delta_{1,\text{sim}} = 0.8$. The bias is shown for varying meta-analysis study sizes, heterogeneity levels, selection type and an increasing selection weight in the estimation.}
\label{fig:App_Bias}
\includegraphics[width = 1\linewidth]{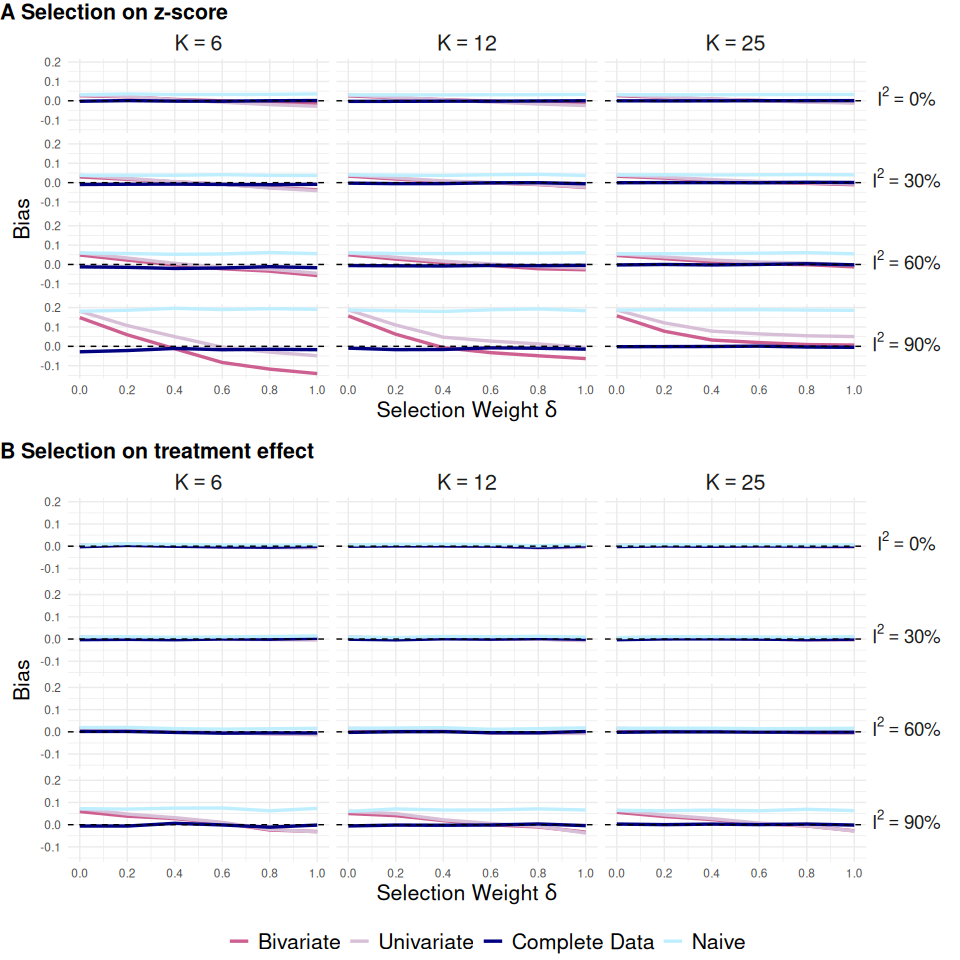}
\end{figure}

The naive estimate exhibits the largest bias across nearly all scenarios (see Figure \ref{fig:App_Bias}). For both the univariate and bivariate ORB-adjusted estimates, bias generally increases as the assumed selection parameter $\delta_{1,\text{est}}$ deviates from the generating value $\delta_{1,\text{sim}}=0.8$, reflecting the impact of misspecifying the reporting mechanism. This effect is most pronounced under high heterogeneity ($I^2=90\%$). As in the main simulation results, bias is generally larger under selection on the z-score than under selection on the treatment effect. Variations in the number of studies have comparatively little influence on the magnitude of bias.

\begin{figure}[!ht]
\centering
\caption{Comparison of estimates for $\theta_1 = \theta_2 = 0.4$, $\rho_B = \rho_W = 0.4$, $p_i = 0.2$ and $\delta_{1,\text{sim}} = 0.8$. The MSE is shown for varying meta-analysis study sizes, heterogeneity levels, selection type and an increasing selection weight in the estimation.}
\label{fig:App_MSE}
\includegraphics[width = 1\linewidth]{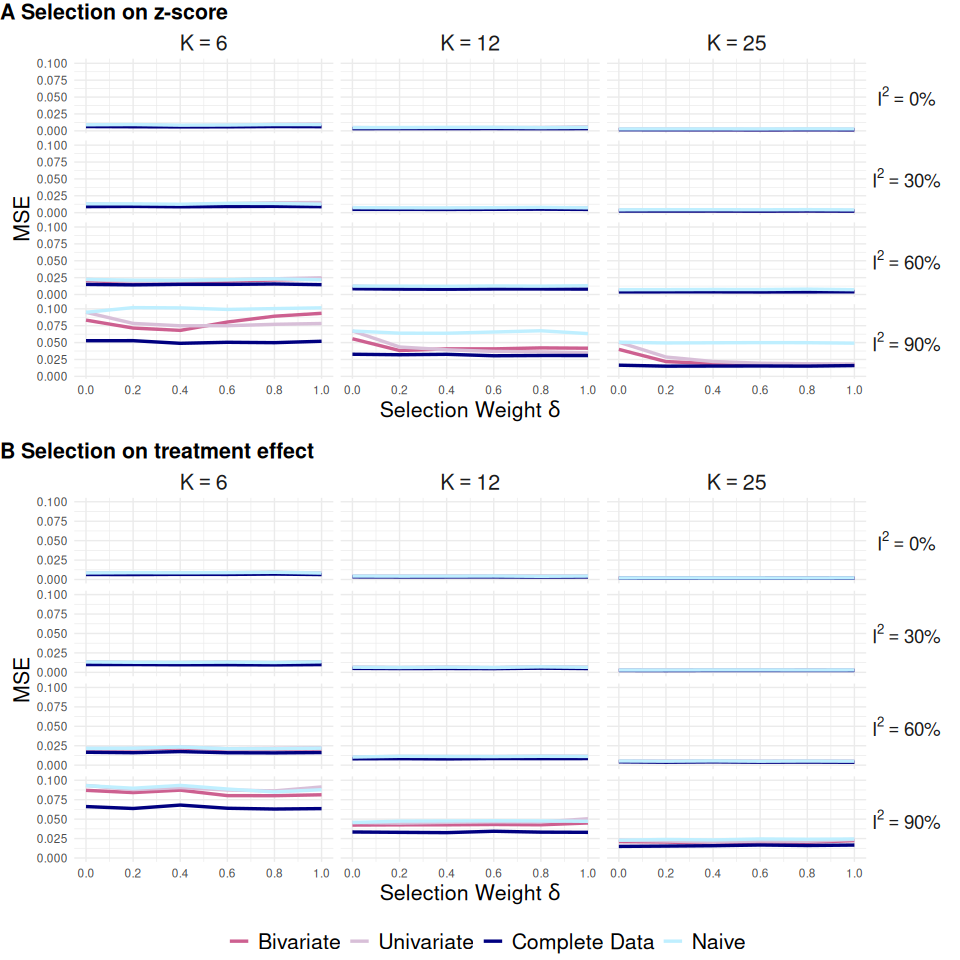}
\end{figure}

The MSE is also highest for the naive estimate and decreases over increasing sample size (see Figure \ref{fig:App_MSE}). Although both adjusted estimates exhibit increased MSE under stronger misspecification, their MSE remains substantially lower than that of the naive estimate. Differences between the univariate and bivariate ORB-adjusted estimates are generally small, indicating that both approaches remain reasonably robust to moderate misspecification of the selection parameter.

\begin{figure}[!ht]
\centering
\caption{Comparison of estimates for $\theta_1 = \theta_2 = 0.4$, $\rho_B = \rho_W = 0.4$, $p_i = 0.2$ and $\delta_{1,\text{sim}} = 0.8$. The CI width is shown for varying meta-analysis study sizes, heterogeneity levels, selection type and an increasing selection weight in the estimation.}
\label{fig:App_CI}
\includegraphics[width = 1\linewidth]{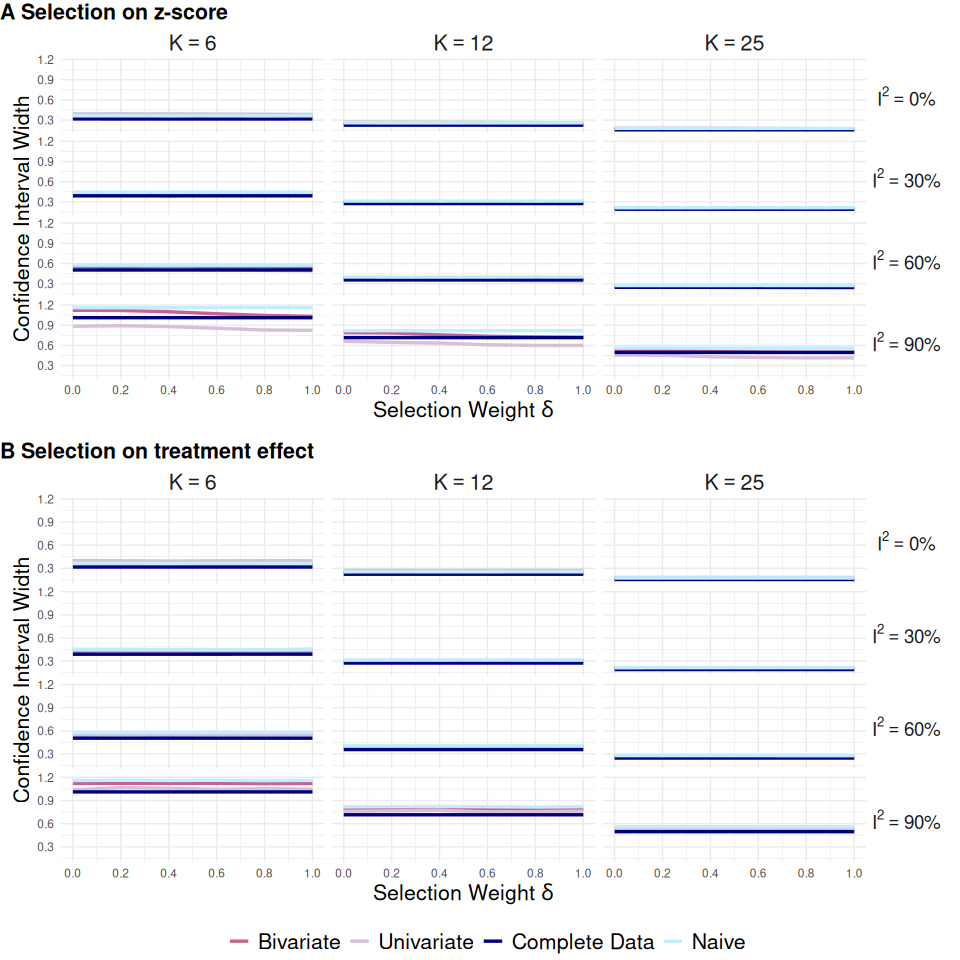}
\end{figure}

CI widths decrease as the number of studies increases, reflecting improved estimation precision. For all scenarios except for high heterogeneity ($I^2 = 90\%$), the complete data estimate has the lowest CI width. For high heterogeneity ($I^2 = 90\%$) the univariate ORB-adjusted estimate has the lowest CI width, reflecting differences in the weighting and imputation procedures under severe selective reporting (see Figure \ref{fig:App_CI}).

\begin{figure}[!ht]
\centering
\caption{Comparison of estimates for $\theta_1 = \theta_2 = 0.4$, $\rho_B = \rho_W = 0.4$, $p_i = 0.2$ and $\delta_{1,\text{sim}} = 0.8$. The coverage is shown for varying meta-analysis study sizes, heterogeneity levels, selection type and an increasing selection weight in the estimation.}
\label{fig:App_Cov}
\includegraphics[width = 1\linewidth]{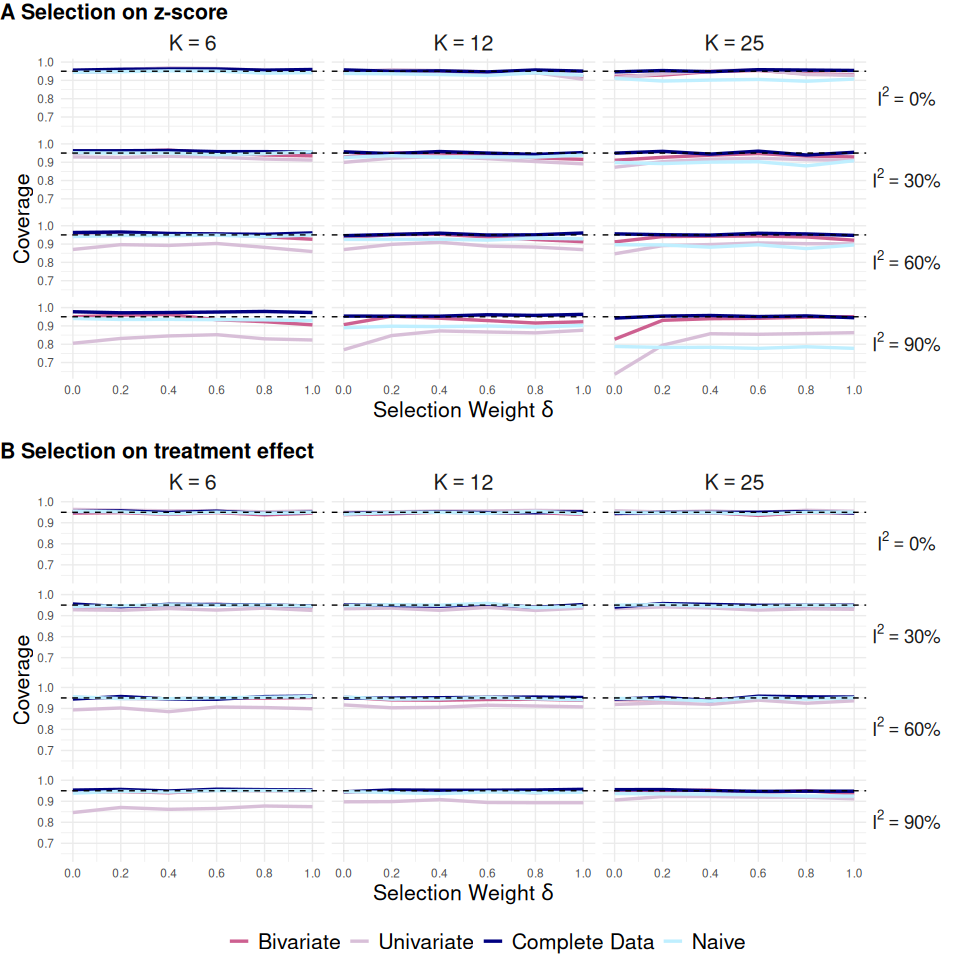}
\end{figure}

As expected, the complete estimate maintains also for model misspecification nominal coverage at approximately 95\% across all simulated scenarios, serving as a reference baseline. The naive estimate shows acceptable nominal coverage when heterogeneity is low to moderate ($I^2 \le 60\%$). However, under strong heterogeneity ($I^2 = 90\%$), its performance degrades as the estimation selection weight increases. Particularly for large sample sizes ($K = 25$) under $z$-score selection coverage drops below 70\% (see Figure \ref{fig:App_Cov} A). The bivariate ORB-adjusted estimate demonstrates robust recovery of the nominal coverage probability across most scenarios. Under extreme heterogeneity ($I^2 = 90\%$), it exhibits a slight drop in coverage at lower estimation selection weights ($\delta < 0.4$) when the estimation model undercorrects for bias. However, it quickly converges toward nominal 95\% coverage as the estimated selection weight matches or approaches the generating parameter ($\delta_{\text{sim}} = 0.8$). The univariate adjusted method underperforms compared to the bivariate model, particularly under high heterogeneity ($I^2 \ge 60\%$) and when the estimated selection weight is low (see Figure \ref{fig:App_Cov}). This under-coverage may arise from a combination of residual variance underestimation and reduced borrowing of strength compared to the bivariate approach.

Overall, the additional analyses demonstrate that both ORB-adjusted methods have substantial advantages over the naive estimate even when the selection mechanism is misspecified. Although bias, MSE, and coverage deteriorate as the assumed selection parameter moves further away from the true generating value, the adjusted estimates generally continue to outperform the naive approach. In particular, the bivariate ORB-adjusted method remains comparatively robust across the considered misspecification scenarios, maintaining lower bias and MSE together with coverage closer to the nominal level. 

%% file: references.bib
@book{agresti2013categorical,
  title={Categorical Data Analysis},
  author={Agresti, Alan},
  year={2002},
  publisher={John Wiley \& Sons}
}

@book{cooper2019handbook,
  title={The Handbook of Research Synthesis and Meta-Analysis},
  author={Cooper, Harris and Hedges, Larry V and Valentine, Jeffrey C},
  year={2019},
  publisher={Russell Sage Foundation}
}

@book{egger2022systematic,
  title={Systematic Reviews in Health Care: Meta-Analysis in Context},
  author={Egger, Matthias and Julian P.T. Higgins and Smith, George Davey},
  year={2022},
  edition={3rd},
  publisher={John Wiley \& Sons}
}

@book{held2020likelihood,
  title={Likelihood and Bayesian Inference},
  author={Held, Leonhard and Bov{\'e}, Daniel Saban{\'e}s},
  journal={Statistics for Biology and Health. Springer, Berlin, Heidelberg},
  year={2020},
  publisher={Springer}
}

@book{schmid2020handbook,
  title={Handbook of Meta-Analysis},
  author={Schmid, Christopher H and Stijnen, Theo and White, Ian},
  year={2020},
  publisher={CRC Press}
}

@article{bai2021bayesian,
  title={{A Bayesian Selection Model for Correcting Outcome Reporting Bias With Application to a Meta-analysis on Heart Failure Interventions}},
  author={Bai, Ray and Liu, Xiaokang and Lin, Lifeng and Liu, Yulun and Kimmel, Stephen E and Chu, Haitao and Chen, Yong},
  journal={arXiv preprint arXiv:2110.08849},
  year={2021}}

@article{beurden2021selective,
  title={{Selective Reporting of Outcomes in Tinnitus Trials: Comparison of Trial Registries With Corresponding Publications}},
  author={Beurden, Isabeau van and Beek, Megan J van de and Heteren, Jan AA van and Smit, Adriana L and Stegeman, Inge},
  journal={Front Neurol},
  volume={12},
  pages={669501},
  year={2021},
  publisher={Frontiers Media SA}}

@article{bresnahan2019topiramate,
  title={{Topiramate Add-on Therapy for Drug-resistant Focal Epilepsy}},
  author={Bresnahan, Rebecca and Hounsome, Juliet and Jette, Nathalie and Hutton, Jane L and Marson, Anthony G},
  journal={Cochrane Database Syst Rev},
  number={10},
  year={2019},
  publisher={John Wiley \& Sons, Ltd}}

@article{butcher2020outcome,
  title={{Outcome Reporting Recommendations for Clinical Trial Protocols and Reports: A Scoping Review}},
  author={Butcher, Nancy J and Mew, Emma J and Monsour, Andrea and Chan, An-Wen and Moher, David and Offringa, Martin},
  journal={Trials},
  volume={21},
  pages={1--17},
  year={2020},
  publisher={Springer}}

@article{chan2004outcome,
  title={{Outcome Reporting Bias in Randomized Trials funded by the Canadian Institutes of Health Research}},
  author={Chan, An-Wen and Krle{\v{z}}a-Jeri{\'c}, Karmela and Schmid, Isabelle and Altman, Douglas G},
  journal={CMAJ},
  volume={171},
  number={7},
  pages={735--740},
  year={2004},
  publisher={CMAJ}}

@article{chan2005identifying,
  title={{Identifying Outcome Reporting Bias in Randomised Trials on PubMed: Review of Publications and Survey of Authors}},
  author={Chan, An-Wen and Altman, Douglas G},
  journal={BMJ},
  volume={330},
  number={7494},
  pages={753},
  year={2005},
  publisher={British Medical Journal Publishing Group}}

@article{chan2014increasing,
  title={{Increasing Value and Reducing Waste: Addressing Inaccessible Research}},
  author={Chan, An-Wen and Song, Fujian and Vickers, Andrew and Jefferson, Tom and Dickersin, Kay and G{\o}tzsche, Peter C and Krumholz, Harlan M and Ghersi, Davina and Van Der Worp, H Bart},
  journal={Lancet},
  volume={383},
  number={9913},
  pages={257--266},
  year={2014},
  publisher={Elsevier}}

@article{chan2017association,
  title={{Association of Trial Registration with Reporting of Primary Outcomes in Protocols and Publications}},
  author={Chan, An-Wen and Pello, Annukka and Kitchen, Jessica and Axentiev, Anna and Virtanen, Jorma I and Liu, Annie and Hemminki, Elina},
  journal={JAMA},
  volume={318},
  number={17},
  pages={1709--1711},
  year={2017},
  publisher={American Medical Association}
}

@article{copas2014model,
  title={{A Model-based Correction for Outcome Reporting Bias in Meta-analysis}},
  author={Copas, John and Dwan, Kerry and Kirkham, Jamie and Williamson, Paula},
  journal={Biostatistics},
  volume={15},
  number={2},
  pages={370--383},
  year={2014},
  publisher={Oxford University Press}}

@article{copas2019model,
  title={{Model-based Sensitivity Analysis for Outcome Reporting Bias in the Meta-analysis of Benefit and Harm Outcomes}},
  author={Copas, John and Marson, Anthony and Williamson, Paula and Kirkham, Jamie},
  journal={Stat Methods Med Res},
  volume={28},
  number={3},
  pages={889--903},
  year={2019},
  publisher={SAGE Publications Sage UK: London, England}}

@article{carpenter2007sensitivity,
  title={{Sensitivity Analysis after Multiple Imputation Under Missing At Random: A Weighting Approach}},
  author={Carpenter, James R and Kenward, Michael G and White, Ian R},
  journal={{Stat Methods Med Res}},
  volume={16},
  number={3},
  pages={259--275},
  year={2007},
  publisher={Sage Publications Sage UK: London, England}
}

@article{carpenter2011assessing,
  title={{Assessing the Sensitivity of Meta-analysis to Selection Bias: A Multiple Imputation Approach}},
  author={Carpenter, James and R{\"u}cker, Gerta and Schwarzer, Guido},
  journal={Biometrics},
  volume={67},
  number={3},
  pages={1066--1072},
  year={2011},
  publisher={Oxford University Press}}

@article{dwan2010assessing,
  title={{Assessing the Potential for Outcome Reporting Bias in a Review: A Tutorial}},
  author={Dwan, Kerry and Gamble, Carrol and Kolamunnage-Dona, Ruwanthi and Mohammed, Shabana and Powell, Colin and Williamson, Paula R},
  journal={Trials},
  volume={11},
  pages={1--10},
  year={2010},
  publisher={Springer}}

@article{dwan2013systematic,
  title={{Systematic Review of the Empirical Evidence of Study Publication Bias and Outcome Reporting Bias—An Updated Review}},
  author={Dwan, Kerry and Gamble, Carrol and Williamson, Paula R and Kirkham, Jamie J and Reporting Bias Group},
  journal={PLoS One},
  volume={8},
  number={7},
  pages={e66844},
  year={2013},
  publisher={Public Library of Science San Francisco, USA}}

@article{fleming2015outcome,
  title={{Outcome Discrepancies and Selective Reporting: Impacting the Leading Journals?}},
  author={Fleming, Padhraig S and Koletsi, Despina and Dwan, Kerry and Pandis, Nikolaos},
  journal={PLoS One},
  volume={10},
  number={5},
  pages={e0127495},
  year={2015},
  publisher={Public Library of Science San Francisco, CA USA}}

@article{frosi2015multivariate,
  title={{Multivariate Meta-analysis helps examine the Impact of Outcome Reporting Bias in Cochrane Rheumatoid Arthritis Reviews}},
  author={Frosi, Giacomo and Riley, Richard D and Williamson, Paula R and Kirkham, Jamie J},
  journal={ J Clin Epidemiol},
  volume={68},
  number={5},
  pages={542--550},
  year={2015},
  publisher={Elsevier}}

@article{glasziou2014reducing,
  title={{Reducing Waste from Incomplete or Unusable Reports of Biomedical Research}},
  author={Glasziou, Paul and Altman, Douglas G and Bossuyt, Patrick and Boutron, Isabelle and Clarke, Mike and Julious, Steven and Michie, Susan and Moher, David and Wager, Elizabeth},
  journal={Lancet},
  volume={383},
  number={9913},
  pages={267--276},
  year={2014},
  publisher={Elsevier}}

@article{hart2012effect,
  title={{Effect of Reporting Bias on Meta-analyses of Drug Trials: Reanalysis of Meta-analyses}},
  author={Hart, Beth and Lundh, Andreas and Bero, Lisa},
  journal={BMJ},
  volume={344},
  pages={d7202},
  year={2012},
  publisher={British Medical Journal Publishing Group}
}

@article{hasselblad1998meta,
  title={{Meta-analysis of Multitreatment Studies}},
  author={Hasselblad, Vic},
  journal={Med Decis Making},
  volume={18},
  number={1},
  pages={37--43},
  year={1998},
  publisher={Sage Publications Sage CA: Thousand Oaks, CA}}

@article{hayati2015evaluation,
  title={{Evaluation of a Weighting Approach for Performing Sensitivity Analysis after Multiple Imputation}},
  author={Hayati Rezvan, Panteha and White, Ian R and Lee, Katherine J and Carlin, John B and Simpson, Julie A},
  journal={BMC Med Res Methodol},
  volume={15},
  number={1},
  pages={83},
  year={2015},
  publisher={Springer}
}

@article{higgins2002quantifying,
  title={{Quantifying Heterogeneity in a Meta-analysis}},
  author={Higgins, Julian PT and Thompson, Simon G},
  journal={Stat Med},
  volume={21},
  number={11},
  pages={1539--1558},
  year={2002},
  publisher={Wiley Online Library}}

@article{hedges1984estimation,
  title={{Estimation of Effect Size under Nonrandom Sampling: The Effects of Censoring Studies yielding Statistically Insignificant Mean Differences}},
  author={Hedges, Larry V},
  journal={J Educ Behav Stat},
  volume={9},
  number={1},
  pages={61--85},
  year={1984},
  publisher={Sage Publications Sage CA: Thousand Oaks, CA}}

@article{hedges1992modeling,
  title={{Modeling Publication Selection Effects in Meta-analysis}},
  author={Hedges, Larry V},
  journal={Statist. Sci.},
  volume={7},
  number={2},
  pages={246--255},
  year={1992},
  publisher={Institute of Mathematical Statistics}}

@article{higgins2009re,
  title={{A Re-evaluation of Random-effects Meta-analysis}},
  author={Higgins, Julian PT and Thompson, Simon G and Spiegelhalter, David J},
  journal={J R Stat Soc Ser A Stat Soc},
  volume={172},
  number={1},
  pages={137--159},
  year={2009},
  publisher={Oxford University Press}}

@article{howard2017systematic,
  title={{Systematic Review: Outcome Reporting Bias is a Problem in High Impact Factor Neurology Journals}},
  author={Howard, Benjamin and Scott, Jared T and Blubaugh, Mark and Roepke, Brie and Scheckel, Caleb and Vassar, Matt},
  journal={PLoS One},
  volume={12},
  number={7},
  pages={e0180986},
  year={2017},
  publisher={Public Library of Science San Francisco, CA USA}}

@article{hwang2018multivariate,
  title={{Multivariate Network Meta-analysis to Mitigate the Effects of Outcome Reporting Bias}},
  author={Hwang, Hyunsoo and DeSantis, Stacia M},
  journal={Stat Med},
  volume={37},
  number={22},
  pages={3254--3266},
  year={2018},
  publisher={Wiley Online Library}}

@article{inthout2014hartung,
  title={{The Hartung-Knapp-Sidik-Jonkman Method for Random Effects Meta-analysis is straightforward and considerably outperforms the Standard DerSimonian-Laird Method}},
  author={IntHout, Joanna and Ioannidis, John PA and Borm, George F},
  journal={BMC Med Res Methodol},
  volume={14},
  number={1},
  pages={25},
  year={2014},
  publisher={Springer}
}

@article{ioannidis2014clinical,
  title={{Clinical Trials: What A Waste}},
  author={Ioannidis, John PA},
  journal={BMJ},
  volume={349},
  year={2014},
  publisher={British Medical Journal Publishing Group}}

@article{jackson2011multivariate,
  title={{Multivariate Meta-analysis: Potential and Promise}},
  author={Jackson, Dan and Riley, Richard and White, Ian R},
  journal={Stat Med},
  volume={30},
  number={20},
  pages={2481--2498},
  year={2011},
  publisher={Wiley Online Library}}

@article{jones2015comparison,
  title={{Comparison of Registered and Published Outcomes in Randomized Controlled Trials: A Systematic Review}},
  author={Jones, Christopher W and Keil, Lukas G and Holland, Wesley C and Caughey, Melissa C and Platts-Mills, Timothy F},
  journal={BMC Med},
  volume={13},
  pages={1--12},
  year={2015},
  publisher={Springer}}

@article{kirkham2010impact,
  title={{The Impact of Outcome Reporting Bias in Randomised Controlled Trials on a Cohort of Systematic Reviews}},
  author={Kirkham, Jamie J and Dwan, Kerry M and Altman, Douglas G and Gamble, Carrol and Dodd, Susanna and Smyth, Rebecca and Williamson, Paula R},
  journal={BMJ},
  volume={340},
  year={2010},
  publisher={British Medical Journal Publishing Group}}

@article{kirkham2012multivariate,
  title={{A Multivariate Meta-Analysis Approach for Reducing the Impact of Outcome Reporting Bias in Systematic Reviews}},
  author={Kirkham, Jamie J and Riley, Richard D and Williamson, Paula R},
  journal={Stat Med},
  volume={31},
  number={20},
  pages={2179--2195},
  year={2012},
  publisher={Wiley Online Library}}

@article{komukai2024publication,
  title={{Publication Bias and Selective Outcome Reporting in Randomized Controlled Trials Related to Rehabilitation: A Literature Review}},
  author={Komukai, Kanako and Sugita, Sho and Fujimoto, Shuhei},
  journal={J Phys Med Rehabil},
  volume={105},
  number={1},
  pages={150--156},
  year={2024},
  publisher={Elsevier}}

@article{lancee2017outcome,
  title={{Outcome Reporting Bias in Randomized-controlled Trials Investigating Antipsychotic Drugs}},
  author={Lancee, M and Lemmens, CMC and Kahn, RS and Vinkers, CH and Luykx, JJ},
  journal={Transl Psychiatry},
  volume={7},
  number={9},
  pages={e1232--e1232},
  year={2017},
  publisher={Nature Publishing Group}}

@article{lancee2022selective,
  title={{Selective Outcome Reporting Across Psychopharmacotherapy Randomized Controlled Trials}},
  author={Lancee, Michelle and Schuring, Marleen and Tijdink, Joeri K and Chan, An-Wen and Vinkers, Christiaan H and Luykx, Jurjen J},
  journal={Int J Methods Psychiatr Res},
  volume={31},
  number={1},
  pages={e1900},
  year={2022},
  publisher={Wiley Online Library}}

@article{langan2019comparison,
  title={{A Comparison of Heterogeneity Variance Estimators in Simulated Random-effects Meta-analyses}},
  author={Langan, Dean and Higgins, Julian PT and Jackson, Dan and Bowden, Jack and Veroniki, Areti Angeliki and Kontopantelis, Evangelos and Viechtbauer, Wolfgang and Simmonds, Mark},
  journal={Res Synth Methods},
  volume={10},
  number={1},
  pages={83--98},
  year={2019},
  publisher={Wiley Online Library}}

@article{lemmens2024outcome,
  title={{Outcome Reporting Bias in Clinical Trials Researching Disease-Modifying Therapy in Patients With Multiple Sclerosis}},
  author={Lemmens, Cynthia MC and van Amerongen, Suzan and Strijbis, Eva M and Killestein, Joep},
  journal={Neurology},
  volume={102},
  number={6},
  pages={e208032},
  year={2024},
  publisher={AAN Enterprises}}

@article{liu2018bayesian,
  title={{Bayesian Mixed Treatment Comparisons Meta-Analysis for Correlated Outcomes Subject to Reporting Bias}},
  author={Liu, Yulun and DeSantis, Stacia M and Chen, Yong},
  journal={J R Stat Soc Ser C Appl Stat},
  volume={67},
  number={1},
  pages={127--144},
  year={2018},
  publisher={Oxford University Press}}

@article{littell2023protocol,
  title={{Protocol: Assessment of Outcome Reporting Bias in Studies included in Campbell Syst Rev}},
  author={Littell, Julia H and Gorman, Dennis M and Valentine, Jeffrey C and Pigott, Therese D},
  journal={Campbell Syst Rev},
  volume={19},
  number={2},
  pages={e1332},
  year={2023},
  publisher={Wiley Online Library}
}

@article{matvienko2024selective,
  title={{Selective Outcome Reporting in Trials of Behavioural Health Interventions in Health Psychology and Behavioural Medicine Journals: A Review}},
  author={Matvienko-Sikar, Karen and O'Shea, Jen and Kennedy, Stephen and Thomas, Siobhan D and Avery, Kerry and Byrne, Molly and McHugh, Sheena and O’Connor, Daryl B and Saldanha, Ian J and Smith, Valerie and others},
  journal={Health Psychol Rev},
  pages={1--15},
  year={2024},
  publisher={Taylor \& Francis}}

@article{mavridis2013practical,
  title={{A Practical Introduction to Multivariate Meta-analysis}},
  author={Mavridis, Dimitris and Salanti, Georgia},
  journal={Stat Methods Med Res},
  volume={22},
  number={2},
  pages={133--158},
  year={2013},
  publisher={SAGE Publications Sage UK: London, England}}

@article{mavridis2014selection,
  title={{A Selection Model for Accounting for Publication Bias in a Full Network Meta-analysis}},
  author={Mavridis, Dimitris and Welton, Nicky J and Sutton, Alex and Salanti, Georgia},
  journal={Stat Med},
  volume={33},
  number={30},
  pages={5399--5412},
  year={2014},
  publisher={Wiley Online Library}
}

@article{mills2013demystifying,
  title={{Demystifying Trial Networks and Network Meta-analysis}},
  author={Mills, Edward J and Thorlund, Kristian and Ioannidis, John PA},
  journal={BMJ},
  volume={346},
  year={2013},
  publisher={British Medical Journal Publishing Group}}

@article{milette2011transparency,
  title={{Transparency of Outcome Reporting and Trial Registration of Randomized Controlled Trials in Top Psychosomatic and Behavioral Health Journals: A Systematic Review}},
  author={Milette, Katherine and Roseman, Michelle and Thombs, Brett D},
  journal={J Psychosom Res},
  volume={70},
  number={3},
  pages={205--217},
  year={2011},
  publisher={Elsevier}}

@article{page2013many,
  title={{Many Scenarios exist for Selective Inclusion and Reporting of Results in Randomized Trials and Systematic Reviews}},
  author={Page, Matthew J and McKenzie, Joanne E and Forbes, Andrew},
  journal={ J Clin Epidemiol},
  volume={66},
  number={5},
  pages={524--537},
  year={2013},
  publisher={Elsevier}}

@article{page2014bias,
  title={{Bias due to Selective Inclusion and Reporting of Outcomes and Analyses in Systematic Reviews of Randomised Trials of Healthcare Interventions}},
  author={Page, Matthew J and McKenzie, Joanne E and Kirkham, Jamie and Dwan, Kerry and Kramer, Sharon and Green, Sally and Forbes, Andrew},
  journal={Cochrane Database Syst Rev},
  number={10},
  year={2014},
  publisher={John Wiley \& Sons, Ltd}}

@article{page2023rob,
  title={{ROB-ME: A Tool for Assessing Risk of Bias due to Missing Evidence in Systematic Reviews with Meta-analysis}},
  author={Page, Matthew J and Sterne, Jonathan AC and Boutron, Isabelle and Hr{\'o}bjartsson, Asbj{\o}rn and Kirkham, Jamie J and Li, Tianjing and Lundh, Andreas and Mayo-Wilson, Evan and McKenzie, Joanne E and Stewart, Lesley A and others},
  journal={BMJ},
  volume={383},
  year={2023},
  publisher={British Medical Journal Publishing Group}}

@article{pulman2014topiramate,
  title={{Topiramate Add-on for Drug-resistant Partial Epilepsy}},
  author={Pulman, Jennifer and Jette, Nathalie and Dykeman, Jonathan and Hemming, Karla and Hutton, Jane L and Marson, Anthony G},
  journal={Cochrane Database Syst Rev},
  number={2},
  year={2014},
  publisher={John Wiley \& Sons, Ltd}}

@article{riley2007evaluation,
  title={{An Evaluation of Bivariate Random-effects Meta-analysis for the Joint Synthesis of Two Correlated Outcomes}},
  author={Riley, Richard D and Abrams, KR and Lambert, PC and Sutton, AJ and Thompson, JR},
  journal={Stat Med},
  volume={26},
  number={1},
  pages={78--97},
  year={2007},
  publisher={Wiley Online Library}}

@article{riley2009multivariate,
  title={{Multivariate Meta-Analysis: The Effect of Ignoring Within-Study Correlation}},
  author={Riley, Richard D},
  journal={J R Stat Soc Ser A Stat Soc},
  volume={172},
  number={4},
  pages={789--811},
  year={2009},
  publisher={Oxford University Press}}

@article{saini2014selective,
  title={{Selective Reporting Bias of Harm Outcomes within Studies: Findings from a Cohort of Systematic Reviews}},
  author={Saini, Pooja and Loke, Yoon K and Gamble, Carrol and Altman, Douglas G and Williamson, Paula R and Kirkham, Jamie J},
  journal={BMJ},
  volume={349},
  year={2014},
  publisher={British Medical Journal Publishing Group}}

@article{saracini2025addressing,
  title={{Addressing Outcome Reporting Bias in Meta-Analysis: A Selection Model Perspective}},
  author={Saracini, Alessandra Gaia and Held, Leonhard},
  journal={Stat Med},
  volume={44},
  number={28-30},
  pages={e70238},
  year={2025},
  publisher={Wiley Online Library}
}

@article{silva2024many,
  title={{Many Randomized Trials in a Large Systematic Review were not registered and had Evidence of Selective Outcome Reporting: A Meta-epidemiological Study}},
  author={Silva, Samuel and Singh, Sareen and Kashif, Shazia and Ogilvie, Rachel and Pinto, Rafael Z and Hayden, Jill A},
  journal={J Clin Epidemiol},
  pages={111568},
  year={2024},
  publisher={Elsevier}
}

@article{shinohara2015protocol,
  title={{Protocol Registration and Selective Outcome Reporting in recent Psychiatry Trials: New Antidepressants and Cognitive Behavioural Therapies}},
  author={Shinohara, Kiyomi and Tajika, Aran and Imai, Hissei and Takeshima, Nozomi and Hayasaka, Yu and Furukawa, Toshi A},
  journal={Acta Psychiatr Scand},
  volume={132},
  number={6},
  pages={489--498},
  year={2015},
  publisher={Wiley Online Library}}

@article{souza2023selective,
  title={{Selective Outcome Reporting Bias is Highly Prevalent in Randomized Clinical Trials of Nonsurgical Periodontal Therapy}},
  author={Souza, Nathalia V and Nicolini, Alessandra C and Dos Reis, Isabella N R and Sendyk, Daniel I and Cavagni, Juliano and Pannuti, Claudio M},
  journal={J Periodontal Res},
  volume={58},
  number={1},
  pages={1--11},
  year={2023},
  publisher={Wiley Online Library}}

@article{song2010dissemination,
  title={{Dissemination and Publication of Research Findings: An updated Review of Related Biases}},
  author={Song, Fujian and Parekh, Sheetal and Hooper, Lee and Loke, Yoon K and Ryder, Jon and Sutton, Alex J and Hing, Caroline and Kwok, Chun Shing and Pang, Chun and Harvey, Ian},
  journal={Health Technol Assess},
  volume={14},
  number={8},
  pages={1--193},
  year={2010}}

@article{sutton2000modelling,
  title={{Modelling Publication Bias in Meta-analysis: A Review}},
  author={Sutton, Alexander J and Song, Fujian and Gilbody, Simon M and Abrams, Keith R},
  journal={Stat Methods Med Res},
  volume={9},
  number={5},
  pages={421--445},
  year={2000},
  publisher={Sage Publications Sage CA: Thousand Oaks, CA}}

@article{tabandeh2022review,
  title={{A Review and Assessment of Importance Sampling Methods for Reliability Analysis}},
  author={Tabandeh, Armin and Jia, Gaofeng and Gardoni, Paolo},
  journal={Structural Safety},
  volume={97},
  pages={102216},
  year={2022},
  publisher={Elsevier}}

@article{thomas2022catalogue,
  title={{Catalogue of Bias: Selective Outcome Reporting Bias}},
  author={Thomas, Elizabeth T and Heneghan, Carl},
  journal={BMJ Evid Based Med},
  volume={27},
  number={6},
  pages={370--372},
  year={2022},
  publisher={Royal Society of Medicine}}

@article{van2024correcting,
  title={{Correcting for Outcome Reporting Bias in a Meta-analysis: A Meta-regression Approach}},
  author={van Aert, Robbie CM and Wicherts, Jelte M},
  journal={Behav Res Methods},
  volume={56},
  number={3},
  pages={1994--2012},
  year={2024},
  publisher={Springer}}

@article{wang2023has,
  title={{Has the Degree of Outcome Reporting Bias in Surgical Randomized Trials changed? A Meta-regression Analysis}},
  author={Wang, Andy and Menon, Rahul and Li, Tom and Harris, Laura and Harris, Ian A and Naylor, Justine and Adie, Sam},
  journal={ANZ J Surg},
  volume={93},
  number={1-2},
  pages={76--82},
  year={2023},
  publisher={Wiley Online Library}}

@article{ward2022outcome,
  title={{Outcome Reporting Bias in Nephrology Randomized Clinical Trials: Examining Outcomes represented by Graphical Illustrations}},
  author={Ward, Frank and Shiely, Frances},
  journal={Contemp Clin Trials Commun},
  volume={28},
  pages={100924},
  year={2022},
  publisher={Elsevier}}

@article{williamson2005identification,
  title={{Identification and Impact of Outcome Selection Bias in Meta-analysis}},
  author={Williamson, Paula R and Gamble, Carrol},
  journal={Stat Med},
  volume={24},
  number={10},
  pages={1547--1561},
  year={2005},
  publisher={Wiley Online Library}}

@article{williamson2005outcome,
  title={{Outcome Selection Bias in Meta-analysis}},
  author={Williamson, Paula R and Gamble, Carrol and Altman, Douglas G and Hutton, JL},
  journal={Stat Methods Med Res},
  volume={14},
  number={5},
  pages={515--524},
  year={2005},
  publisher={Sage Publications Sage CA: Thousand Oaks, CA}}

@article{williamson2007application,
  title={{Application and Investigation of a Bound for Outcome Reporting Bias}},
  author={Williamson, Paula R and Gamble, Carrol},
  journal={Trials},
  volume={8},
  pages={1--12},
  year={2007},
  publisher={Springer}}

@article{zhang2025threat,
  title={{The Threat of Serious Outcome Reporting Bias in Randomized Controlled Trials on Acute Ischemic Stroke to Evidence Synthesis: A Meta-epidemiological Study}},
  author={Zhang, Na and Long, Youlin and Wang, Xinyao and Wang, Xinyi and Guo, Qiong and Li, Zhengchi and Du, Liang},
  journal={Cardiovasc Diagn Ther},
  volume={15},
  number={6},
  pages={1182--1193},
  year={2025},
  publisher={LWW}
}
